\documentclass[conference,a4paper
]{IEEEtran}
\usepackage[dvips]{graphicx,color}
\usepackage{latexsym}
\usepackage{amssymb}
\usepackage{amsmath,bm}

\usepackage{array}

\begin{document}

\makeatletter
\@dblfptop 0pt
\makeatother

%
%
%
%
\newcommand{\qed}{\hfill$\square$}
\newcommand{\suchthat}{\mbox{~s.t.~}}
\newcommand{\markov}{\leftrightarrow}
%
%
\newenvironment{pRoof}{%
 \noindent{\em Proof.\ }}{%
 \hspace*{\fill}\qed \\
 \vspace{2ex}}


\newcommand{\ket}[1]{| #1 \rangle}
\newcommand{\bra}[1]{\langle #1 |}
\newcommand{\bol}[1]{\mathbf{#1}}
\newcommand{\rom}[1]{\mathrm{#1}}
\newcommand{\san}[1]{\mathsf{#1}}
\newcommand{\mymid}{:~}
\newcommand{\argmax}{\mathop{\rm argmax}\limits}
\newcommand{\argmin}{\mathop{\rm argmin}\limits}
%
%
%
%
\newcommand{\bc}{\begin{center}}  %
\newcommand{\ec}{\end{center}}
\newcommand{\befi}{\begin{figure}[h]}  %
\newcommand{\enfi}{\end{figure}}
\newcommand{\bsb}{\begin{shadebox}\begin{center}}   %
\newcommand{\esb}{\end{center}\end{shadebox}}
\newcommand{\bs}{\begin{screen}}     %
\newcommand{\es}{\end{screen}}
\newcommand{\bib}{\begin{itembox}}   %
\newcommand{\eib}{\end{itembox}}
\newcommand{\bit}{\begin{itemize}}   %
\newcommand{\eit}{\end{itemize}}
\newcommand{\defeq}{:=}
\newcommand{\Qed}{\hbox{\rule[-2pt]{3pt}{6pt}}}
\newcommand{\beq}{\begin{equation}}
\newcommand{\eeq}{\end{equation}}
\newcommand{\beqa}{\begin{eqnarray}}
\newcommand{\eeqa}{\end{eqnarray}}
\newcommand{\beqno}{\begin{eqnarray*}}
\newcommand{\eeqno}{\end{eqnarray*}}
\newcommand{\ba}{\begin{array}}
\newcommand{\ea}{\end{array}}
\newcommand{\vc}[1]{\mbox{\boldmath $#1$}}
\newcommand{\lvc}[1]{\mbox{\scriptsize \boldmath $#1$}}
\newcommand{\svc}[1]{\mbox{\scriptsize\boldmath $#1$}}

\newcommand{\wh}{\widehat}
\newcommand{\wt}{\widetilde}
\newcommand{\ts}{\textstyle}
\newcommand{\ds}{\displaystyle}
\newcommand{\scs}{\scriptstyle}
\newcommand{\vep}{\varepsilon}
\newcommand{\rhp}{\rightharpoonup}
\newcommand{\cl}{\circ\!\!\!\!\!-}
\newcommand{\bcs}{\dot{\,}.\dot{\,}}
\newcommand{\eqv}{\Leftrightarrow}
\newcommand{\leqv}{\Longleftrightarrow}
\newtheorem{co}{Corollary} 
\newtheorem{lm}{Lemma} 
\newtheorem{Ex}{Example} 
\newtheorem{Th}{Theorem}
\newtheorem{df}{Definition} 
\newtheorem{pr}{Property} 
\newtheorem{pro}{Proposition} 
\newtheorem{rem}{Remark} 

\newcommand{\lcv}{convex } 

\newcommand{\hugel}{{\arraycolsep 0mm
                    \left\{\ba{l}{\,}\\{\,}\ea\right.\!\!}}
\newcommand{\Hugel}{{\arraycolsep 0mm
                    \left\{\ba{l}{\,}\\{\,}\\{\,}\ea\right.\!\!}}
\newcommand{\HUgel}{{\arraycolsep 0mm
                    \left\{\ba{l}{\,}\\{\,}\\{\,}\vspace{-1mm}
                    \\{\,}\ea\right.\!\!}}
\newcommand{\huger}{{\arraycolsep 0mm
                    \left.\ba{l}{\,}\\{\,}\ea\!\!\right\}}}

\newcommand{\Huger}{{\arraycolsep 0mm
                    \left.\ba{l}{\,}\\{\,}\\{\,}\ea\!\!\right\}}}

\newcommand{\HUger}{{\arraycolsep 0mm
                    \left.\ba{l}{\,}\\{\,}\\{\,}\vspace{-1mm}
                    \\{\,}\ea\!\!\right\}}}

\newcommand{\hugebl}{{\arraycolsep 0mm
                    \left[\ba{l}{\,}\\{\,}\ea\right.\!\!}}
\newcommand{\Hugebl}{{\arraycolsep 0mm
                    \left[\ba{l}{\,}\\{\,}\\{\,}\ea\right.\!\!}}
\newcommand{\HUgebl}{{\arraycolsep 0mm
                    \left[\ba{l}{\,}\\{\,}\\{\,}\vspace{-1mm}
                    \\{\,}\ea\right.\!\!}}
\newcommand{\hugebr}{{\arraycolsep 0mm
                    \left.\ba{l}{\,}\\{\,}\ea\!\!\right]}}
\newcommand{\Hugebr}{{\arraycolsep 0mm
                    \left.\ba{l}{\,}\\{\,}\\{\,}\ea\!\!\right]}}

\newcommand{\HugebrB}{{\arraycolsep 0mm
                    \left.\ba{l}{\,}\\{\,}\vspace*{-1mm}\\{\,}\ea\!\!\right]}}

\newcommand{\HUgebr}{{\arraycolsep 0mm
                    \left.\ba{l}{\,}\\{\,}\\{\,}\vspace{-1mm}
                    \\{\,}\ea\!\!\right]}}

\newcommand{\hugecl}{{\arraycolsep 0mm
                    \left(\ba{l}{\,}\\{\,}\ea\right.\!\!}}
\newcommand{\Hugecl}{{\arraycolsep 0mm
                    \left(\ba{l}{\,}\\{\,}\\{\,}\ea\right.\!\!}}
\newcommand{\HUgecl}{{\arraycolsep 0mm
                    \left(\ba{l}{\,}\\{\,}\\{\,}\vspace{-1mm}
                    \\{\,}\ea\right.\!\!}}

\newcommand{\hugecr}{{\arraycolsep 0mm
                    \left.\ba{l}{\,}\\{\,}\ea\!\!\right)}}
\newcommand{\Hugecr}{{\arraycolsep 0mm
                    \left.\ba{l}{\,}\\{\,}\\{\,}\ea\!\!\right)}}
\newcommand{\HUgecr}{{\arraycolsep 0mm
                    \left.\ba{l}{\,}\\{\,}\\{\,}\vspace{-1mm}
                    \\{\,}\ea\!\!\right)}}

\newcommand{\hugepl}{{\arraycolsep 0mm
                    \left|\ba{l}{\,}\\{\,}\ea\right.\!\!}}
\newcommand{\Hugepl}{{\arraycolsep 0mm
                    \left|\ba{l}{\,}\\{\,}\\{\,}\ea\right.\!\!}}
\newcommand{\hugepr}{{\arraycolsep 0mm
                    \left.\ba{l}{\,}\\{\,}\ea\!\!\right|}}
\newcommand{\Hugepr}{{\arraycolsep 0mm
                    \left.\ba{l}{\,}\\{\,}\\{\,}\ea\!\!\right|}}

\newcommand{\MEq}[1]{\stackrel{
{\rm (#1)}}{=}}

\newcommand{\MLeq}[1]{\stackrel{
{\rm (#1)}}{\leq}}

\newcommand{\ML}[1]{\stackrel{
{\rm (#1)}}{<}}

\newcommand{\MGeq}[1]{\stackrel{
{\rm (#1)}}{\geq}}

\newcommand{\MG}[1]{\stackrel{
{\rm (#1)}}{>}}

\newcommand{\MPreq}[1]{\stackrel{
{\rm (#1)}}{\preceq}}

\newcommand{\MSueq}[1]{\stackrel{
{\rm (#1)}}{\succeq}}

\newcommand{\Ch}{{\Gamma}}
\newcommand{\Rw}{{W}}

\newcommand{\Cd}{{\cal R}_{\rm d}(\Ch)}
\newcommand{\CdB}{{\cal R}_{\rm d}^{\prime}(\Ch)}
\newcommand{\CdBB}{{\cal R}_{\rm d}^{\prime\prime}(\Ch)}

\newcommand{\Cdi}{{\cal R}_{\rm d}^{\rm (in)}(\Ch)}
\newcommand{\Cdo}{{\cal R}_{\rm d}^{\rm (out)}(\Ch)}

\newcommand{\tCdi}{\tilde{\cal R}_{\rm d}^{\rm (in)}(\Ch)}
\newcommand{\tCdo}{\tilde{\cal R}_{\rm d}^{\rm (out)}(\Ch)}
\newcommand{\hCdo}{  \hat{\cal R}_{\rm d}^{\rm (out)}(\Ch)}

\newcommand{\Cs}{{\cal R}_{\rm s}(\Ch)}
\newcommand{\CsB}{{\cal R}_{\rm s}^{\prime}(\Ch)}
\newcommand{\CsBB}{{\cal R}_{\rm s}^{\prime\prime}(\Ch)}

\newcommand{\Csi}{{\cal R}_{\rm s}^{\rm (in)}(\Ch)}
\newcommand{\Cso}{{\cal R}_{\rm s}^{\rm (out)}(\Ch)}
\newcommand{\tCsi}{\tilde{\cal R}_{\rm s}^{\rm (in)}(\Ch)}
\newcommand{\tCso}{\tilde{\cal R}_{\rm s}^{\rm (out)}(\Ch)}
\newcommand{\cCsi}{\check{\cal R}_{\rm s}^{\rm (in)}(\Ch)}
\newcommand{\Cds}{{\cal C}_{\rm ds}(\Ch)}
\newcommand{\Cdsi}{{\cal C}_{\rm ds}^{\rm (in)}(\Ch)}
\newcommand{\Cdso}{{\cal C}_{\rm ds}^{\rm (out)}(\Ch)}
\newcommand{\tCdsi}{\tilde{\cal C}_{\rm ds}^{\rm (in)}(\Ch)}
\newcommand{\tCdso}{\tilde{\cal C}_{\rm ds}^{\rm (out)}(\Ch)}
\newcommand{\hCdso}{\hat{\cal C}_{\rm ds}^{\rm (out)}(\Ch)}
\newcommand{\Css}{{\cal C}_{\rm ss}(\Ch)}
\newcommand{\Cssi}{{\cal C}_{\rm ss}^{\rm (in)}(\Ch)}
\newcommand{\Csso}{{\cal C}_{\rm ss}^{\rm (out)}(\Ch)}
\newcommand{\tCssi}{\tilde{\cal C}_{\rm ss}^{\rm (in)}(\Ch)}
\newcommand{\tCsso}{\tilde{\cal C}_{\rm ss}^{\rm (out)}(\Ch)}
\newcommand{\Cde}{{\cal R}_{\rm d1e}(\Ch)}
\newcommand{\Cdei}{{\cal R}_{\rm d1e}^{\rm (in)}(\Ch)}
\newcommand{\Cdeo}{{\cal R}_{\rm d1e}^{\rm (out)}(\Ch)}
\newcommand{\tCdei}{\tilde{\cal R}_{\rm d1e}^{\rm (in)}(\Ch)}
\newcommand{\tCdeo}{\tilde{\cal R}_{\rm d1e}^{\rm (out)}(\Ch)}
\newcommand{\hCdeo}{  \hat{\cal R}_{\rm d1e}^{\rm (out)}(\Ch)} 
\newcommand{\Cse}{{\cal R}_{\rm s1e}(\Ch)}
\newcommand{\Csei}{{\cal R}_{\rm s1e}^{\rm (in)}(\Ch)}
\newcommand{\Cseo}{{\cal R}_{\rm s1e}^{\rm (out)}(\Ch)}
\newcommand{\tCsei}{\tilde{\cal R}_{\rm s1e}^{\rm (in)}(\Ch)}
\newcommand{\tCseo}{\tilde{\cal R}_{\rm s1e}^{\rm (out)}(\Ch)}

\newcommand{\Capa}{C}

\newcommand{\ZeTa}{\zeta(S;Y,Z|U)}
\newcommand{\ZeTaI}{\zeta(S_i;Y_i,Z_i|U_i)}

\newcommand{\CEreg}{\irBr{rate} }
\newcommand{\CEregB}{rate\MarkOh{-equivocation }}

\newcommand{\Cls}{class NL}
\newcommand{\vSpa}{\vspace{0.3mm}}
\newcommand{\Prmt}{\zeta}
\newcommand{\pj}{\omega_n}

\newfont{\bg}{cmr10 scaled \magstep4}
\newcommand{\bigzerol}{\smash{\hbox{\bg 0}}}
\newcommand{\bigzerou}{\smash{\lower1.7ex\hbox{\bg 0}}}
\newcommand{\nbn}{\frac{1}{n}}
\newcommand{\ra}{\rightarrow}
\newcommand{\la}{\leftarrow}
\newcommand{\ldo}{\ldots}
\newcommand{\ep}{\epsilon }
\newcommand{\typi}{A_{\epsilon }^{n}}
\newcommand{\bx}{\hspace*{\fill}$\Box$}
\newcommand{\pa}{\vert}
\newcommand{\ignore}[1]{}

\newcommand{\dBar}{||}
\newcommand{\One}{\rm (i)}
\newcommand{\Two}{\rm (ii)}
\newcommand{\Thr}{\rm (iii)}
\newcommand{\Fou}{\rm (iv)}
\newcommand{\Fiv}{\rm (v)}
\newcommand{\Six}{\rm (vi)}
\newcommand{\SDC}{\rm SDC}
\newcommand{\GPC}{\rm GPC}

%
%
%
%

\newenvironment{jenumerate}
	{\begin{enumerate}\itemsep=-0.25em \parindent=1zw}{\end{enumerate}}
\newenvironment{jdescription}
	{\begin{description}\itemsep=-0.25em \parindent=1zw}{\end{description}}
\newenvironment{jitemize}
	{\begin{itemize}\itemsep=-0.25em \parindent=1zw}{\end{itemize}}
\renewcommand{\labelitemii}{$\cdot$}

\newcommand{\iro}[2]{{\color[named]{#1}#2\normalcolor}}
\newcommand{\irr}[1]{{\color[named]{Black}#1\normalcolor}}

\newcommand{\irg}[1]{{\color[named]{Green}#1\normalcolor}}
\newcommand{\irb}[1]{{\color[named]{Black}#1\normalcolor}}

\newcommand{\irBl}[1]{{\color[named]{Black}#1\normalcolor}}
\newcommand{\irWh}[1]{{\color[named]{White}#1\normalcolor}}

\newcommand{\irY}[1]{{\color[named]{Yellow}#1\normalcolor}}
\newcommand{\irO}[1]{{\color[named]{Orange}#1\normalcolor}}
\newcommand{\irBr}[1]{{\color[named]{Black}#1\normalcolor}}

\newcommand{\IrBr}[1]{{\color[named]{Purple}#1\normalcolor}}
\newcommand{\irBw}[1]{{\color[named]{Brown}#1\normalcolor}}
\newcommand{\irPk}[1]{{\color[named]{Magenta}#1\normalcolor}}
\newcommand{\irCb}[1]{{\color[named]{CadetBlue}#1\normalcolor}}

%
\newenvironment{indention}[1]{\par
\addtolength{\leftskip}{#1}\begingroup}{\endgroup\par}
%
\newcommand{\namelistlabel}[1]{\mbox{#1}\hfill} 
\newenvironment{namelist}[1]{%
\begin{list}{}
{\let\makelabel\namelistlabel
\settowidth{\labelwidth}{#1}
\setlength{\leftmargin}{1.1\labelwidth}}
}{%
\end{list}}
%
%
\newcommand{\bfig}{\begin{figure}[t]}
\newcommand{\efig}{\end{figure}}
\setcounter{page}{1}

\newtheorem{theorem}{Theorem}
\newcommand{\Ep}{\mbox{\rm e}}

\newcommand{\Exp}{\exp
}
\newcommand{\idenc}{{\varphi}_n}
\newcommand{\resenc}{
{\varphi}_n}
\newcommand{\ID}{\mbox{\scriptsize ID}}
\newcommand{\TR}{\mbox{\scriptsize TR}}
\newcommand{\Av}{\mbox{\sf E}}

\newcommand{\Vl}{|}
\newcommand{\Ag}{(R,P_{X^n}|W^n)}
\newcommand{\Agv}[1]{({#1},P_{X^n}|W^n)}
\newcommand{\Avw}[1]{({#1}|W^n)}

\newcommand{\Jd}{X^nY^n}
\newcommand{\IdR}{r_n}

\newcommand{\Index}{{n,i}}

\newcommand{\cid}{C_{\mbox{\scriptsize ID}}}
\newcommand{\cida}{C_{\mbox{{\scriptsize ID,a}}}}

\newcommand{\BiBArXiv}{
\bibitem{GPStConvArXiv16}
Y. Oohama, ``Exponent function for state dependent channels 
with  full state information at the sender and partial 
state Information at the receiver at rates above 
the capacity,'' {\it preprint;} available at 
http://arxiv.org/PS\_cache.

}
\newcommand{\ArXiv}{in \cite{GPStConvArXiv16}. }
\newcommand{\GorF}{found }
\newcommand{\GorFb}{is found }
\newcommand{\Comment}{}
\newcommand{\Delete}{\irr{\mbox{``The equation has been deleted.''}}} 
\newcommand{\Vincent}[1]{{\color[named]{Black}#1\normalcolor}}

\newcommand{\pOne}{\alpha}
\newcommand{\pTwo}{\beta}
\newcommand{\pThr}{\lambda}

\arraycolsep 0.5mm
\date{}
%
\title{
%
Strong Converse Exponent for State Dependent Channels 
With  Full State Information at the Sender 
}
\author{%
\IEEEauthorblockN{Yasutada Oohama}
\IEEEauthorblockA{
  University of Electro-Communications, Tokyo, Japan \\
  Email: oohama@uec.ac.jp} 
} 
\newcommand{\Empty}{

\author{%
Yasutada Oohama 
\thanks{
Y. Oohama is with 
University of Electro-Communications,
1-5-1 Chofugaoka Chofu-shi, Tokyo 182-8585, Japan.
}%
\thanks{
}
}
\markboth{
}
{
}
}

\maketitle

\begin{abstract}
We consider the state dependent channels with full state 
information with at the sender. For this state dependent 
channel, the channel capacity was determined by Gel'fand and 
Pinsker. In this paper, we study the correct probability of decoding at 
rates above the capacity. We prove that when the transmission rate is 
above the capacity this probability goes to zero exponentially 
and derive an explicit lower bound of this exponent function. 
\end{abstract}

\section{Introduction}

In this paper we study the classical problem of channel coding with 
noncausal state information at the encoder. 
This problem was posed and investigated by 
the Gel'fand and Pinsker \cite{gelfand:80}. 
We hereafter call the above state dependent channel 
the Gel'fand and Pinsker channel (the GPC).
The channel capacity of the GPC was determined by 
Gel'fand and Pinsker \cite{gelfand:80}. 
Tyagi and Narayan \cite{tyagi:09} 
proved the strong converse theorem for the GPC.

In this paper, we establish the exponential strong converse 
theorem for the GPC. We study the correct probability of decoding 
at rates above the capacity and prove that when the transmission 
rate is above the capacity this probability goes to 
zero exponentially and derive an explicit lower 
bound of this exponent function. 

We derive the result by a combination of two methods. 
One is a method called the recursive method. This method includes 
a certain recursive algorithm for a single letterization of exponent 
functions. The recursive method is a general powerful tool to prove 
strong converse theorems for several coding problems in information 
theory. In fact, this method plays important roles in deriving exponential 
strong converse exponent for communication systems treated 
in \cite{oohama2015DBCexponent}, 
\cite{oohama2015DBCFBex},
\cite{oohama:18}, 
\cite{oohama:19},
\cite{oohama20ABCStConv}.
The other is to use the  information spectrum version of Csisz\'ar's 
sum identity found in \cite{csiszar:78}. For cooperation with the recursive 
method, some careful use of the identity is necessary. 
Similar combination of the two methods to prove the 
exponential strong converse theorem is found 
in \cite{oohama20ABCStConv}.

The  state dependent channel with full state information at the sender 
and partial state information at the receiver (the SDC) was posed and 
investigated by Heegard and El Gamal \cite{heegard:83}. The SDC can be 
regarded an extension of the GPC. Steineberg \cite{steinberg:08} 
determined the capacity region for the SDC which indicates the trade off 
between the transmission rate and the compression rate on the partial 
state information. 

The exponential strong converse theorem for the SDC was first 
investigated by the author \cite{oohamaItw16}. However, we have a gap on 
the proof. In this paper we fix this gap in the special case that the 
state information is not available at the decoder.   

\section{Coding Problem for State Dependent Channels}

%
%

In the channel coding with noncausal state information at the encoder
we would like to send a 
uniformly distributed message over a state-dependent channel 
$W^n:{\cal X}^n \times S^n$, where ${\cal S}, {\cal X}$ 
and ${\cal Y}$, respectively, are the state, input and output alphabets.

\begin{figure}[t]
\bc
\includegraphics[width=5.5cm]{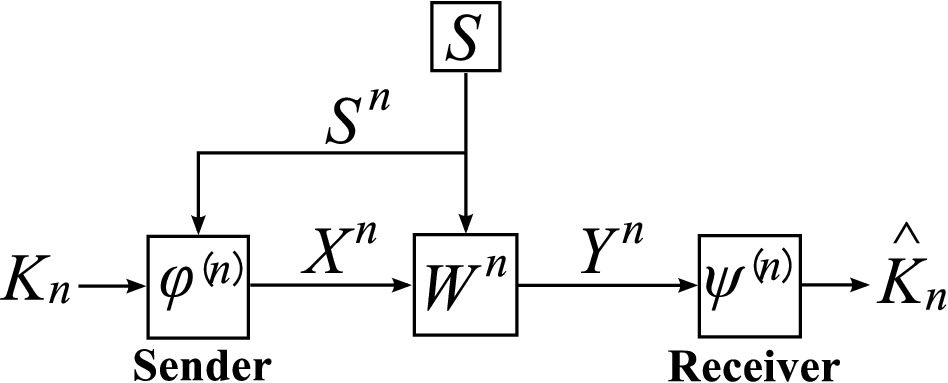}
\caption{Coding for state dependent channels}
\label{fig:GP} 
\ec
\vspace*{-5mm}
\end{figure}

We assume that ${\cal X}, {\cal Y},$ ${\cal S}$ are finite sets.
The state-dependent channel(SDC) we study in this paper is defined 
by a stationary discrete memoryless channel specified by the following 
stochastic matrix:
\beq
W \defeq \{W(y|x,s)\}_{
(s,x,y) 
\in    {\cal S}
\times {\cal X} 
\times {\cal Y}}.
\eeq
Let $X^n$ be a random variable taking values in 
${\cal X}^n$. We write an element of ${\cal X}^n$ as   
$x^n=x_{1}x_{2}$$\cdots x_{n}.$ 
Suppose that $X^n$ has a probability distribution on ${\cal X}^n$ 
denoted by 
$p_{X^n}=$ 
$\left\{p_{X^n}(x^n) 
\right\}_{{x^n} \in {\cal X}^n}$.
Similar notations are adopted for other random variables. 
Let $Y^n \in {\cal Y}^n$ be a random variable 
obtained as the channel output by connecting 
$X^n$ to the input of channel under the random state $S^n$. 
We write a conditional distribution of $Y^n$ on ${\cal Y}^n$ 
given $X^n$ and $S^n$ 
as 
$$
W^n=\left\{
W^n(y^n|x^n,s^n)\right
\}_{(s^n,x^n,y^n) \in {\cal S}^n \times {\cal X}^n \times {\cal Y}^n}.
$$
Since the channel is memoryless, we have 
\beq
W^n({y}^n|x^n,s^n)=\prod_{t=1}^nW (y_t|x_t,s_t).
\label{eqn:sde0}
\eeq
We assume that the state information of $S^n$ is an output of 
a stationary discrete memoryless source $\{S_t\}_{t=1}^{\infty}$ 
specified with a probability distribution $p_S=\{p_S(s)\}_{s \in {\cal S}}$
on ${\cal S}$. 
Transmission of messages via the state dependent channel 
is shown in Fig. \ref{fig:GP}. 
The random variable $K_n$ is a message sent to the receiver.
The random variable $S_n$ represent a random state.
Under $S^n$, a sender transforms $K_n$ into a transmitted 
sequence $X^n$ using an encoder function $\varphi^{(n)}$ 
and sends it to the receiver. 
In this paper we assume that the 
encoder function $\varphi^{(n)}$ is a stochastic encoder. 
In this case, $\varphi^{(n)}$ is a stochastic matrix given by
$$
\varphi^{(n)}=\{
\varphi^{(n)}(x^n|k,s^n)\}_{
(k,s^n,x^n)\in {\cal K}_n \times {\cal S}^n \times {\cal X}^n},
$$ 
where $\varphi^{(n)}(x^n|k,s^n)$ is a conditional probability 
of $x^n \in {\cal X}^n$ given $k\in$ ${\cal K}_n$ 
and non-causal random state $s^n\in {\cal S}^n$.
The joint probability mass function on 
${\cal K}_n $ 
$\times {\cal S}^n$ 
$\times {\cal X}^n$ 
$\times {\cal Y}^n$ 
is given by
\begin{align*}
& \Pr\{(K_n,
S^n,X^n,Y^n)=(k,
s^n,x^n, y^n)\}
\nonumber\\
&= \frac{1}{\pa{\cal K}_n\pa}
\varphi^{(n)}(x^n|k,s^n)p_{S^n}(s^n)
\prod_{t=1}^n W\left(y_t \left|x_t,s_t\right.\right),
\end{align*}
where $\pa {\cal K}_n \pa$ is a cardinality 
of the set ${\cal K}_n$. The decoding function 
at the receiver is denoted by ${\psi}^{(n)}$. 
This function is formally defined by
$
{\psi}^{(n)}: {\cal Y}^{n} \to {\cal K}_n.
$
The average error probability of decoding on the receiver 
is defined by 
\beqno
& &{\rm P}_{\rm e}^{(n)}=
   {\rm P}_{\rm e}^{(n)}(\varphi^{(n)},\psi^{(n)})
\defeq 
\Pr\{\psi^{(n)}(Y^n)\neq K_n\}.
\eeqno
For $k\in {\cal K}_n$, set
$
{\cal D}(k)\defeq \{ y^n: \psi^{(n)}(y^n)=(k) \}.
$
A family of sets $\{{\cal D}(k) \}_{k \in {\cal K}_n}$ 
is called the decoding region. Using the decoding region, 
${\rm P}_{\rm e}^{(n)}$ can be written as
\beqno
{\rm P}_{\rm e}^{(n)}
&=&\frac{1}{|{\cal K}_n|} 
\sum_{k \in {\cal K}_n}
\sum_{\scs (s^n,x^n,y^n)
\in {\cal S}^n \times {\cal X}^n\times {\cal Y}^n:
       \atop{
       \scs y^n   \in {\cal D}^{c}(k)
       }
    }  
\\
& &\times 
\varphi^{(n)}(x^n|k,s^n)W^n(y^n|x^n,s^n)p_{S^n}(s^n).
\eeqno
Set 
\beqno
& &{\rm P}^{(n)}_{\rm c}=
   {\rm P}^{(n)}_{\rm c}(\varphi^{(n)},\psi^{(n)})
\defeq 
1-{\rm P}^{(n)}_{\rm e}(\varphi^{(n)}, \psi^{(n)}).
\eeqno
The quantity ${\rm P}^{(n)}_{\rm c}$ is called the average 
correct probability of decoding. This quantity has the 
following form:
\beqno
{\rm P}_{\rm c}^{(n)}
&=&\frac{1}{|{\cal K}_n|} 
\sum_{k \in {\cal K}_n }
\sum_{\scs (s^n,x^n,y^n)\in 
{\cal S}^n \times {\cal X}^n \times {\cal Y}^n:
       \atop{
       \scs y^n \in {\cal D}(k)
       }
    }  
\\
& &\times 
\varphi^{(n)}(x^n|k,s^n)W^n(y^n|x^n,s^n)p_{S^n}(s^n).
\eeqno
For fixed $\varepsilon \in (0,1)$, $R$ 
is $\varepsilon$-{\it achievable} if there exists 
a sequence of pairs 
$\{(\varphi^{(n)},$ 
$ \psi^{(n)})\}_{n=1}^{\infty}$ 
such that  for any $\delta>0$ and for any $n$ with 
$n\geq n_0=n_0(\varepsilon,\delta)$,
\beqa 
{\rm P}_{\rm e}^{(n)}&\leq &\varepsilon, 
\nbn \log \pa {\cal K}_n \pa \geq  R-\delta.
\nonumber
\eeqa
The supremum of all $\varepsilon$-achievable rates is denoted by 
$C_{\GPC}($ $\varepsilon|W)$, which is called 
the $\varepsilon$-capacity of the Gel'fand-Pinsker 
channel (GPC). 
We set
$$
C_{\GPC}(p_S,W)
\defeq \inf_{\varepsilon\in(0,1)}
C_{\GPC}(\varepsilon|p_S,W),
$$
which is called the capacity of the GPC. 
We can show that the capacity regions 
$C_{\rm \GPC }(\varepsilon|$ $p_S,W)$, 
$\varepsilon \in (0,1)$, and 
$ C_{\rm \GPC}(p_S,W)$
satisfy the following property. 
\begin{pr}\label{pr:pro0a}
$\quad$
%
$C_{\rm \GPC}(\varepsilon|p_S,W)$ has another 
form using $(n,\vep)$-capacity 
${C}_{\rm \GPC}(n,\varepsilon |p_S,W)$,
the definition of which is as follows.
We set 
\beqno
& &{C}_{\rm \GPC}(n,\varepsilon |p_S,W)
 \defeq \sup \{R: 
\\
& &\mbox{ There exists }(\varphi^{(n)},\psi^{(n)})
\mbox{ such that }
\\
& &\frac{1}{n}\log |{\cal K}_n|\leq R,  
{\rm P}_{\rm e}^{(n)}(\varphi^{(n)},\psi^{(n)})
\leq \varepsilon\}.
\eeqno 
Using ${C}_{\rm \GPC}(n,$ $\varepsilon| p_S,W)$, 
${C}_{\rm \GPC}(\varepsilon| p_S,W)$ can be 
expressed as 
\beqno
C_{\rm \GPC}(\varepsilon|p_S,W)
&=&\sup_{m \geq 1}
\inf_{n \geq m}C_{\rm \GPC}(n,\varepsilon|p_S,W).
\eeqno

\end{pr}

Proof of Property \ref{pr:pro0a} is given in Appendix \ref{sub:ApdaAAAA}. 
\newcommand{\ApdaAAAA}{
\subsection{
General Properties on ${C}_{\rm GPC}(n,\varepsilon,\Gamma|W)$.
}
\label{sub:ApdaAAAA}

In this appendix we prove Property \ref{pr:pro0a} describing 
general properties on ${C}_{\rm GPC}(n,\varepsilon|p_S,W)$.

{\it Proof of Property \ref{pr:pro0a}:} 
We first prove the inequality 
$$
{C}_{\rm GPC}(\varepsilon,\Gamma|W)
\geq \sup_{m \geq 1} \underline{C}_{\rm GPC}(m,\varepsilon,\Gamma|W).
$$
We assume that 
$$
R \leq \sup_{m \geq 1} 
\underline{C}_{\GPC}(m,\varepsilon|p_S,W).$$ 
Then, there 
exists positive integer $m$ such that 
$R \leq \underline{C}_{\GPC}(m,$ $\varepsilon|W)$. 
Then, by the definition of 
$\underline{C}_{\rm GPC}(m,\varepsilon|W)$, we have that for any 
$n\geq m$, there exists a pair $(\varphi^{(n)},\psi^{(n)})$ 
such that 
\beq
{\rm P}_{{\rm e}}^{(n)}(\varphi^{(n)},\psi^{(n)}
) \leq \varepsilon, 
\quad \frac{1}{n}\log |{\cal K}_n| \geq R.
\label{eqn: SssdF}
\eeq
It is obvious that under (\ref{eqn: SssdF}), we have 
for any $\delta>0$, and any $n\geq m $, we have 
\beq
{\rm P}_{{\rm e}}^{(n)}(\varphi^{(n)}, \psi^{(n)})
\leq \varepsilon, 
\quad \frac{1}{n}\log |{\cal K}_n| \geq R-\delta.
\label{eqn: SssdFz}
\eeq
The bound (\ref{eqn: SssdFz}) implies that 
$R \leq C_{\rm GPC}(\varepsilon|p_S,W)$. Hence 
the bound 
$$
{C}_{\rm GPC}(\varepsilon,\Gamma|W)\geq 
\sup_{m \geq 1} \underline{C}_{\rm GPC}(m,\varepsilon|p_S,W)
$$ 
is proved. We next prove the reverse inequality. We assume that
$R\leq C_{\rm GPC}(\varepsilon|p_S,W)$. Then there exists 
$\{(\varphi^{(n)}, \psi^{(n)})$ $ \}_{n\geq 1}$ such that
for any $\delta> 0$ and any $n$ with 
$n \geq n_0=n_0(\varepsilon,\delta)$
we have that
\beq
{\rm P}_{{\rm e}}^{(n)}(\varphi^{(n)}, \psi^{(n)}) 
\leq \varepsilon, 
\quad \frac{1}{n}\log |{\cal K}_n| \geq R-\delta.
\label{eqn: SssdFzz}
\eeq
The bound (\ref{eqn: SssdFzz}) implies that 
\beqno 
R-\delta &\leq & 
\underline{C}_{\rm GPC}(n_0,\varepsilon|p_S,W)
\leq \sup_{n \geq 1}\underline{C}_{\rm GPC}(n,\varepsilon|p_S,W).
\eeqno
On the other hand, by the first assumption we have 
$R-\delta\leq C_{\rm GPC}(\varepsilon|p_S,W)-\delta$. 
Hence, we have
$$
C_{\rm GPC}(\varepsilon|p_S,W)-\delta
\leq \sup_{n\geq 1}\underline{C}_{\rm GPC}(n,\varepsilon| p_S,W).
$$
Since we can take $\delta>0$ arbitrary small, we have
$$
C_{\rm GPC}(\varepsilon|p_S,W)
\leq \sup_{n \geq 1}\underline{C}_{\rm GPC}(n,\varepsilon|p_S,W),
$$ 
completing the proof. \hfill \IEEEQED
}
To describe previous works on ${C}_{\GPC}(p_S,W)$ 
and ${C}_{\GPC}(\varepsilon|p_S, W)$, 
we introduce an auxiliary random variable $U$ 
taking values in a finite 
set ${\cal U}$ $\times {\cal V}$. We assume that 
the joint distribution of $(U,S,X,Y)$ is 
\beqno 
& & p_{USXY}(u,s,x,y)
\\
&=&p_{U}(u)p_{SX|UV}(s,x|u,v)W(y|x,s). 
\eeqno
The above condition is equivalent to 
$U\leftrightarrow (X,S) \leftrightarrow Y$.
Define the set of probability 
distributions $q=q_{USXY}$ of $(U,$ $S,$ $X,$ $Y)$ $\in$ 
${\cal U}$ 
$\times{\cal S}$ 
$\times{\cal X}$ 
$\times{\cal Y}$
by
\begin{align*}
{\cal P}(p_S,W)
& \defeq  
\{q: \pa{\cal U}\pa \leq \pa{\cal S}\pa \pa{\cal X} \pa,
q_S=p_S, q_{Y|XS}=W, 
\vSpa\\
& U \markov (X,S) \markov Y \}.
\end{align*}
Set 
\beqno
& &C(p_S,W) \defeq
\max_{q \in {\cal P}(p_S,W)}\{I_q(Y;U)-I_q(S;U)\}.
\eeqno
The GPC was posed and investigated 
by Gel'fand and Pinsker \cite{gelfand:80}. They 
determined $C_{\GPC}(p_S,W)$. Their result is the following: 
\begin{Th}[Gel'fand and Pinsker \cite{gelfand:80}] For any state 
dependent channel $W$,
$C_{\GPC}(p_S,W)=C(p_S,W).$
\end{Th}

The strong converse theorem for the GPC was proved by Tyagi 
and Narayan \cite{tyagi:09}. Their result is the following:
\begin{Th}[Tyagi and Narayan \cite{tyagi:09}]\label{Th:TN09} For each 
$\varepsilon$ $ \in (0,1)$, and for any state dependent 
channel $W$, we have
$$
 {C}_{\GPC}(\varepsilon|p_S,W)={C}_{\GPC}(p_S,W).
$$
\end{Th}


To prove Theorem \ref{Th:TN09}, they used a method of image size 
characterization introduced by 
Csisz\'ar and K\"orner \cite{csiszar-korner:81}. Capacity 
theorems for the state dependent channel in the case of 
general noisy channels was obtained by Tan \cite{tan:13}. 
To derive those  capacity results he used the information 
spectrum method introduced by Han \cite{han:book}. 

To examine an asymptotic behavior of ${\rm P}_{\rm c}^{(n)}$ 
for rates outside the capacity region ${C}(p_S,W)$
we define the following quantity. 
\beqno
& & 
G^{(n)}(R|p_S,W)
\\
&&
\defeq
\min_{\scs 
(\varphi^{(n)},\psi^{(n)}):
    \atop{\scs 
         \atop{\scs 
         (1/n)\log | {\cal K}_n |\geq R
         }
    }
}
\hspace*{-2mm}
\left(-\frac{1}{n}\right)
\log {\rm P}_{\rm c}^{(n)}
(\varphi^{(n)},\psi^{(n)}).
\eeqno
By time sharing we have 
that 
$\{G^{(n)}(\irb{R}|p_S,\irBr{W})\}_{n\geq 1}$ 
satisfies the following subadditivity property:
\begin{align*}
& G^{(n+m)}(\irb{R}|p_S,\irBr{W}) 
\\
& \leq \frac{nG^{(n)}(\irb{R}|p_S,\irBr{W}) 
+m G^{(m)}(\irb{R}|p_S,\irBr{W})}{n+m}.
\end{align*}
Hence we have
\beqno
\lim_{n\to\infty}G^{(n)}(\irb{R}|p_S,\irBr{W}) 
=\inf_{n\geq 1}G^{(n)}(\irb{R}|p_S,\irBr{W}).
\eeqno
Set
\beqno
& &
G(R|p_S,W) \defeq \inf_{n\geq 1}G^{(n)}(R|p_S,W),
\\
& &{\cal R}(p_S,\irBr{W}) \defeq \{(\irb{R},G): 
G\geq G(\irb{R}|p_S, \irBr{W})\}. 
\eeqno
The exponent function $G(R|p_S,W)$ is a convex function 
of $R$. In fact, 
by time sharing we have that 
\begin{align*}
& G^{(n+m)}\left(\left.
\frac{n R +m R^{\prime}}{n+m}
\right|p_S, W \right) 
\\
& \leq \frac{
nG^{(n)}(R|p_S,W)+mG^{(m)}(R^{\prime}|p_S,W)}
{n+m}, 
\end{align*}
from which we have that for any $\kappa \in [0,1]$
\begin{align*}
& G(\kappa R+ \bar{\kappa} R^{\prime}|p_S,W)
\\
&\leq 
\kappa G(R|p_S,W)+ \bar{\kappa} G(R^{\prime}|p_S,W).
\end{align*}


\section{Main Results}


In this section we state our main results. 
We first define several quantities to describe our 
main results. Define the set of probability 
distributions $p=p_{USXY}$ of $(U,$ $S,$ $X,$ $Y)$ $\in$ 
${\cal U}$ $\times{\cal S}$ $\times{\cal X}$ $\times{\cal Y}$ 
by  
\begin{align*}
&{\cal Q}
 \defeq 
 \{q=q_{USXY}:\pa {\cal U} \pa \leq 
    \pa {\cal S} \pa+  \pa {\cal Y} \pa -1\}.
\end{align*}
Furthermore define 
\begin{align*}
& \omega^{(\pOne)}_{q}(s,x,y|u)
\defeq
\log \frac{p_{S}(s)W(y|x,s)}{q_{S}(s)q_{Y|XSU}(y|x,s,u)}
\\
&\quad +\pOne \log \frac{q_{Y|U}(y|u)q_{S}(s)}
                        {q_{S|U}(s|u)q_{Y}(y)},
\\
& \Lambda^{(\pOne,\pTwo)}(q|p_S,W)
\defeq {\rm E}_{q}\left[
\exp\left\{\pTwo \omega^{(\pOne)}_{q}(S,X,Y|U)\right\}
\right],
\\
&            \Omega^{(\pOne,\pTwo)}(q|p_S,W)
\defeq \log \Lambda^{(\pOne,\pTwo)}(q|p_S,W),
\\
&  \Omega^{(\pOne, \pTwo)}(p_S,W)
\defeq \max_{q \in{\cal Q}}
\Omega^{(\pOne,\pTwo)}(q|p_S,W),
\\
& F^{(\pOne,\pTwo)}(R|p_S,W)
\defeq
\frac{ \pTwo \pOne R
       -\Omega^{(\pOne,\pTwo)}(p_S,W)}
        {1+\pTwo[1+3\pOne]}, 
\\
& F(R|p_S,W)\defeq 
\sup_{\scs \pOne,\pTwo \geq 0}
F^{(\pOne,\pTwo)}(R|p_S,W).
\end{align*}
We next define a function serving as a lower bound of 
$F(R|p_S, W)$. For each 
$p=p_{UXYS}\in {\cal P}(p_S,W)$, define
\begin{align*}
& \tilde{\omega}_{p}(s,x,y|u)
\defeq   \log \frac{p_{Y|U}(y|u)p_{S}(s)}
                   {p_{S|U}(s|u)p_{Y}(y)},
\\
& \tilde{\Lambda}^{(\pThr)}(p) \defeq 
{\rm E}_{p} \left[
\exp\left\{\pThr\tilde{\omega}_p(S,X,Y|U)\right\}
\right],
\\
& \tilde{\Omega}^{(\pThr)}(p) \defeq 
\log \tilde{\Lambda}^{(\pThr)}(p).
\end{align*}
Furthermore, set
\begin{align*}
& \tilde{\Omega}^{(\pThr)}(p_S,W)
 \defeq \max_{ p \in {\cal P}(p_S,W)}
 \tilde{\Omega}^{(\pThr)}(p),
\\
&\tilde{F}^{(\pThr)}(R|p_S,W)
\defeq
\frac{\pThr R-\tilde{\Omega}^{(\pThr)}(p_S,W)}
{2+7\pThr},
\\
& \tilde{F}(R|p_S,W)
 \defeq \sup_{\scs  \pThr \geq 0} \tilde{F}^{(\pThr)}(R|p_S,W).
\end{align*}
We can show that the above functions and sets 
satisfy the following property. 
\begin{pr}\label{pr:pro1}  
$\quad$
\begin{itemize}
\item[a)]
The cardinality bound 
$\pa {\cal U} \pa \leq 
 \pa {\cal S} \pa + \pa {\cal Y} \pa-1$
appearing in the definition of ${\cal Q}$ is sufficient 
to describe $\Omega^{(\pOne,\pTwo)}(p_S,W)$.
Furthermore, the cardinality bound 
$  \pa{\cal U}\pa \leq 
$ $\pa{\cal S}\pa \pa{\cal X}\pa $
appearing in the definition of ${\cal P}(p_S,W)$
is sufficient to describe 
$\tilde{\Omega}^{(\pThr)}(p_S,W)$. 

\item[b)] 
For any $R>0$, we have  
\beqno
& &F(R|p_S,W) \geq \tilde{F}(R|p_S,W).
\eeqno

\item[c)]
For any $\lambda \in [0,1/2]$ and any 
$q \in {\cal P}(p_S,W)$, we have 
$$
{|{\cal S}|^{-\frac{1}{2}}}
\leq \tilde{\Lambda}^{(\lambda)}(q)\leq 
|{\cal X}|^2|{\cal S}|^2|{\cal Y}|.
$$


\item[d)]
Fix any $q \in {\cal P}(p_S,W)$. 
For $\lambda \in [0,1]$, we define a probability distribution 
$q^{(\lambda)}=q_{USXY}^{(\lambda)}$ by
\begin{align*} 
& q^{(\lambda)}(u,s,x,y)
\\
&\defeq
\frac{
q(u,s,x,y)
\exp\left\{\lambda
\tilde{\omega}_{q }(s,x,y|u)\right\}
}{
{\rm E}_{q}
\left[\exp\left\{\lambda 
\tilde{\omega}_q(S,X,Y|U)\right\}\right]}.
\end{align*}
\irr{Then, for $\lambda \in [0,1/4]$, $\tilde{\Omega}^{(\lambda)}(q)$ 
is twice differentiable. Furthermore, for 
$\lambda \in [0,1/4]$, we have }
\beq
\left.
\ba{l}
\ds \frac{\rm d}{{\rm d}\lambda} \tilde{\Omega}^{(\lambda)}(q)
={\rm E}_{q^{(\lambda)}}
\left[\tilde{\omega}_q(S,X,Y|U)\right],
\vspace{1mm}\\
\ds  \frac{\rm d^2}{{\rm d}\lambda^2} 
\tilde{\Omega}^{(\lambda)}(q)
={\rm Var}_{q^{(\lambda)}}
\left[\tilde{\omega}_{q}(S,X,Y|U)\right].
\ea
\right\}
\label{eqn:Sddpp}
\eeq

\item[e)]
For \irr{$\lambda\in [0,1]\times [0,1/4]$}, define
\beqno
& &\rho^{(\lambda)}(p_S,W)
\\
& &\defeq
\irr{\max}_{\scs (\irr{\gamma,}q) \in \irr{[0,\lambda]}
      \atop{\scs \irr{\times}{\cal P}(p_S,W):
            \atop{\scs \tilde{\Omega}^{(\lambda)}(q)
                 \atop{
                 \scs =\tilde{\Omega}^{(\lambda)}(p_S,W)
                 }
            }
     }
}
{\rm Var}_{\irr{q^{(\gamma)}}}
\left[\tilde{\omega}_{q}(S,X,Y|U)\right],
\eeqno
and set
\beqno
& &\rho=\rho(p_S,W)
\defeq 
\max_{\scs \lambda \in [0,1/4]}
\rho^{(\lambda)}(p_S,W).
\eeqno
Then, we have $ \rho(p_S,W)< \infty$. Furthermore, 
for any $\lambda \in [0,1/4]$, we have 
$$
\tilde{\Omega}^{(\lambda)}(p_S,W) 
\leq \lambda C(p_S,W) +\frac{\lambda^2}{2}\rho(p_S,W). 
$$
\item[f)] For every $\tau \in (0,(1/4)\rho(p_S,W)]$,
$R > C(p_S,W)+\tau$ 
implies that 
\begin{align*}
& \tilde{F}(R|p_S,W) > \frac{\rho(p_S,W)}{4}\cdot {g}^2
\left(\frac{\tau}{\rho(p_S,W)}\right)>0,
\end{align*}
where $g$ is the inverse function of 
${\vartheta}(a) \defeq a+(7/4)a^2, a\geq 0$.

\end{itemize}

\end{pr}

Proof of Property \ref{pr:pro1} part a) is given in 
Appendix \ref{sub:ApdaAAAb}. Proofs of Property \ref{pr:pro1} 
part b) is given in Appendix \ref{sub:ApdaAACa}. 
Proofs of Property \ref{pr:pro1} parts c), d), e) are 
given in Appendix \ref{sub:ApdaAACb}. 
Proof of Property \ref{pr:pro1} part f) are 
given in Appendix \ref{sub:ApdaAACc}. 
%
\newcommand{\ApdaAAAb}{
\subsection{
Cardinality Bound on Auxiliary Random Variables
}
\label{sub:ApdaAAAb}

In this appendix we prove Property \ref{pr:pro1} part a).

{\it Proof of Property \ref{pr:pro1} part a):} 
We first bound  the cardinality $|{\cal U}|$ of ${U}$ 
to show that the bound 
$|{\cal U}| \leq |{\cal S}|$$+|{\cal Y}|-1$ 
is sufficient to describe 
$\hat{\Omega}_n^{(\pOne,\pTwo)}(p_S,W)$. 
We first observe that 
\begin{align}
& \Lambda^{(\pOne,\pTwo)}(q)
=\sum_{u\in {\cal U}}q_U(u)
\zeta^{(\pOne,\pTwo)}(q_{SXY|U}(\cdot|u),q_S,q_Y),
\label{eqn:Oldaqadz}
\end{align}
where we set
\begin{align*}
& \zeta^{(\pOne,\pTwo)}(q_{SXY|U}(\cdot|u),q_S,q_Y)
\\
&\defeq  \sum_{\scs (s,x,y)\atop{\scs 
\in {\cal S}\times{\cal X}\times{\cal Y}}}
q_{SXY|U}(s,x,y|u)
\exp\left\{\pOne \omega^{
  (\pOne)}_{q}(s,x,y|u)\right\}.
\end{align*}
For each $u \in{\cal U}$, 
$
\zeta^{(\pOne,\pTwo)}$ $(q_{SXY|U}(\cdot|u),q_S,q_Y)
$
is a continuous function of $q_{SXY|U}(\cdot|u)$. 
We further observe that 
\beq
\left. 
\ba{rcl}
q_{S}(s)
&=& \ds \sum_{u\in {\cal U}}q_U(u)q_{S|U}(s|u),
\vSpa\\
q_{Y}(y)
&=& \ds \sum_{u\in {\cal U}}q_U(u)q_{Y|U}(s|u).
\ea
\right\}
\label{eqn:Oldasdftt}
\eeq
Then by the support lemma, 
\beq
  1+ |{\cal U}|\leq |{\cal S}|+|{\cal Y}|-2=|{\cal S}|+|{\cal Y}|-1
\label{eqn:OldQzzA}
\eeq
is sufficient to express one value of (\ref{eqn:Oldaqadz}) and 
$|{\cal S}|+|{\cal Y}|-2$ values of (\ref{eqn:Oldasdftt}). 
We next 
show that the bound 
$|{\cal U}| \leq |{\cal S}||{\cal X}|$
is sufficient to describe 
$\tilde{\Omega}^{(\pThr)}$ $(p_S,W)$. 
Observe that 
\begin{align}
& p_{SX}(s,x)
=\sum_{u\in {\cal U}}q_U(u)
p_{SX|U}(s,x|u),
\label{eqn:asdf}
\\
& \tilde{\Lambda}^{(\pThr)}(p)
=\sum_{u \in {\cal U}}p_U(u)
\tilde{\zeta}^{(\pThr)}(p_{SXY|U}(\cdot|u)),
\label{eqn:aqqqa}
\end{align}
where we set
\begin{align*}
& \tilde{\zeta}^{(\pThr)}(p_{SXY|U}(\cdot|u))
\\
&\defeq  \sum_{\scs (s,x,y)\atop{\scs 
\in {\cal S}\times{\cal X}\times{\cal Y}}}
q_{SXY|U}(s,x,y|u)
\exp\left\{ \pThr \tilde{\omega}_{p}(s,x,y|u)\right\}.
\end{align*}
For each $u \in{\cal U}$, 
$
\tilde{\zeta}^{(\pThr)}$ $(p_{SXY|U}(\cdot|u))
$
is a continuous function of $p_{SXY|U}(\cdot|u)$.
Then by the support lemma, 
\beq
  |{\cal U}| \leq |{\cal S}||{\cal X}|-1+1
= |{\cal S}||{\cal X}|
\label{eqn:QzzA}
\eeq
is sufficient to express $|{\cal S}||{\cal X}|-1$ 
values of (\ref{eqn:asdf}) and one value of (\ref{eqn:aqqqa}).
\hfill \IEEEQED
}
%
%
%
%
%
\newcommand{\ApdaAACa}{
\subsection{
Proof of Property \ref{pr:pro1} part b)
} 
\label{sub:ApdaAACa}

In this appendix we prove Property \ref{pr:pro1} part b).
Fix $q=q_{USXY} \in{\cal Q}$, arbitrary. 
For $p_{USXY}=(p_S,q_{XU|S},p_S,W) \in {\cal P}(p_S,W)$, define
\begin{align*}
& \hat{\omega}_{p,q_{Y|U}}(s,y|u)
\defeq  \log \frac{q_{Y|U}(y|u)}{p_{Y}(y)} 
+ \log \frac{p_{S}(s)}{p_{S|U}(s|u)},
\\
& \hat{\Omega}^{(\pOne)}(p,q_{Y|U})
\defeq 
\log 
{\rm E}_{p}
\left[\exp\left\{\pTwo \hat{\omega}_{p,q_{Y|U}}(S,Y|U)\right\}\right].
\end{align*}

Then we have the following two lemmas.  
\begin{lm}\label{lm:lemmaSS}
For any $\pOne \geq 0$, 
$\pTwo\in [0,\frac{1}{1+2\pOne}]$ and 
any $q=q_{UXYZ}\in {\cal Q}$, there exists
$p=p_{USXY}\in {\cal P}(p_S,W)$ such that 
\beq
{\Omega}^{(\pOne,\pTwo)}(q|p_S,W)
\leq 
\pTwo \hat{\Omega}^{(\pOne)}(p,q_{Y|U}).
\label{eqn:Zsss}
\eeq
\end{lm}

\begin{lm}\label{lm:lemmaSSz}
For any $\pOne$ satisfying  $\pOne \in [0,1)$,  
any $p=p_{UXYZ}\in {\cal P}(p_S,W)$, and any stochastic 
matrix $q_{Y|U}$ induced by $q_{UXYZ}\in {\cal Q}$, we have 
\begin{align}
& \hat{\Omega}^{(\pOne)}(p,q_{Y|U})
\leq  \bar{\pOne} \tilde{\Omega}^{(\frac{\pOne}{\bar{\pOne}})}(p).
\label{eqn:ZsssZZZ}
\end{align}
\end{lm}

From Lemmas \ref{lm:lemmaSS} and \ref{lm:lemmaSSz} we have 
the following corollary.
\begin{co}\label{co:CoOne}
For any $\pOne,\pTwo$ satisfying
$\pOne \in [0,1)$,  
$\pTwo\in [0,\frac{1}{1+2\pOne}]$ and any 
$q=q_{UXYZ}\in {\cal Q}$, there exists 
$p=p_{UXYZ}\in {\cal P}(p_S,W)$
such that 
\begin{align}
& \Omega^{(\pOne,\pTwo)}(q|p_S,W)
\leq \pTwo \bar{\pOne}\tilde{\Omega}^{(\frac{\pOne}{\bar{\pOne}})}(p).
\label{eqn:ZsssZZZqq}
\end{align}
From (\ref{eqn:ZsssZZZqq}), we have that
for any $\pOne \in [0,1)$,  
$\pTwo\in [0,\frac{1}{1+2\pOne}]$, we have 
\begin{align}
& \Omega^{(\pOne,\pTwo)}(p_S,W)
\leq \pTwo\bar{\pOne}
\tilde{\Omega}^{(\frac{\pOne}{\bar{\pOne}})}(p_S,W).
\label{eqn:ZsssAZZqq}
\end{align}
\end{co}

{\it Proof of Lemma \ref{lm:lemmaSS}:} We fix 
$\pOne \geq 0$, $\pTwo \in [0,1]$ arbitrary. For each 
$q=q_{USXY} \in {\cal Q}$, we choose $p=p_{USXY} \in {\cal P}(p_S,W)$ 
so that $p=(p_S,q_{XU|S},W)$. 
Then for any 
$(u,x,y,z)$ 
$\in {\cal U}$
$\times {\cal S}$
$\times {\cal X}$
$\times {\cal Y}$, 
we have the following: 
\begin{align}
& \frac{p_S(s)}{q_{S}(s)} \cdot
  \frac{W(y|x,s)}{q_{Y|XSU}(y|x,s,u)}
 =\frac{p_S(s)q_{XU|S}(x,u|s)}{q_{S}(s)q_{XU|S}(x,u|s)} 
\notag\\
&\quad \times \frac{W(y|x,s)}{q_{Y|XSU}(y|x,s,u)}
=\frac{p_{USXY}(u,s,x,y)}{q_{USXY}(u,s,x,y)}.
\label{eqn:Zsppp}
\end{align}
On upper bounds of $\exp \left\{ 
{\Omega}^{(\pOne,\pTwo)}(q|p_S,W)
\right\}$, we have the following chain of inequalities:
\begin{align}
& \exp \left\{ 
{\Omega}^{(\pOne,\pTwo)}(q|p_S,W)
\right\}
\MEq{a}
{\rm E}_{q}
\Hugebl
\Hugel
\frac{p_{USXY}(U,S,X,Y)}{q_{USXY}(U,S,X,Y)}
\nonumber\\
&\quad \times 
\frac{q_{Y|U}^{\pOne}(Y|U)}
       {p_{Y}^{\pOne}(Y)}
\frac{p_{S}^{\pOne}(S)}
   {p_{S|U}^{\pOne}(S|U)}
\Huger^{\pTwo}
\nonumber\\
&\quad \times 
\left.
\left\{ 
\frac{p_{Y}^{\frac{2\pOne\pTwo}{\bar{\pTwo}}}(Y)}
     {q_{Y}^{\frac{2\pOne\pTwo}{\bar{\pTwo}}}(Y)}
\right\}^{\frac{\bar{\pTwo}}{2}}
\left\{ 
\frac{p_{S|U}^{\frac{2\pOne \pTwo}{\bar{\pTwo}}}(S|U)}
     {q_{S|U}^{\frac{2\pOne\pTwo}{\bar{\pTwo}}}(S|U)}
\right\}^{\frac{\bar{\pTwo}}{2}}
\right]
\nonumber\\
&\MLeq{b}
\exp \left\{ \pTwo
\hat{\Omega}^{(\pOne)}
  (p,q_{Z|U})\right\}
A_1^{\frac{\bar{\pTwo}}{2}}
A_2^{\frac{\bar{\pTwo}}{2}},
\label{eqn:EddxSSS}
\end{align}
where we set
\beqno
A_1 &\defeq& {\rm E}_q\left[
\frac{p_{Y}^{\frac{2\pOne \pTwo}{\bar{\pTwo}}}(Y)}
     {q_{Y}^{\frac{2\pOne \pTwo}{\bar{\pTwo}}}(Y)}
\right],
A_2 \defeq {\rm E}_q\left[
\frac{p_{S|U}^{\frac{2\pOne \pTwo}{\bar{\pTwo}}}(S|U)}
     {q_{S|U}^{\frac{2\pOne \pTwo}{\bar{\pTwo}}}(S|U)}
\right].
\eeqno
Step (a) follows from (\ref{eqn:Zsppp}). Step (b) 
follows from H\"older's inequality. From (\ref{eqn:EddxSSS}), we can 
see that it suffices to show $A_i \leq 1,i=1,2$ to 
complete the proof. Note here that when $\pTwo \in [0,\frac{1}{1+2\pOne}]$,
we have ${2\pOne(\pTwo}/{\bar{\pTwo})}\leq 1$.
Hence under $\pTwo \in [0,\frac{1}{1+2\pOne}]$, we can apply H\"older's 
inequality to $A_1$ to obtain 
\begin{align*}
&A_1={\rm E}_q \left[
\frac{p_{Y}^{\frac{2\pOne\pTwo}{\bar{\pTwo}}}(Y)}
     {q_{Y}^{\frac{2\pOne\pTwo}{\bar{\pTwo}}}(Y)} \right]
\leq \left({\rm E}_q\left[
\frac{p_{Y}(Y)}{q_{Y}(Y)} \right]
     \right)^{\frac{2\pOne\pTwo}{\bar{\pTwo}}}
=1.
\end{align*}
In a similar manner we can prove $A_2\leq 1$. 
Hence we have (\ref{eqn:Zsss}) in Lemma \ref{lm:lemmaSS}. 
\hfill \IEEEQED 

{\it Proof of Lemma \ref{lm:lemmaSSz}:}
We fix $\pOne \in [0,1)$,  
$\pTwo\in [0,\frac{1}{1+2\pOne}]$, arbitrary.
For any 
$p=p_{USXY}\in {\cal P}_{\rm sh}(p_S,W)$, 
and any $q=q_{USXY}\in {\cal Q}$, we have the following 
chain of inequalities: 
\begin{align}
& \exp \left\{ 
\hat{\Omega}^{(\pOne)}(p,q_{Z|U})
\right\}
\nonumber\\
&=
{\rm E}_{p}
\left[
\left\{
\frac{p_{Y|U}^{\frac{\pOne}{\bar{\pOne}}}(Y|U)}
       {p_{Y}^{\frac{\pOne}{\bar{\pOne}}}(Y)}
  \frac{p_{S}^{\frac{\pOne}{\bar{\pOne}}}(S)}
     {p_{S|U}^{\frac{\pOne}{\bar{\pOne}}}(S|U)
}\right\}^{\bar{\pOne}}
\left\{\frac{q_{Y|U}(Y|U)}
            {p_{Y|U}(Y|U)}\right\}^{\pOne}
\right]
\nonumber\\
&\MLeq{a}
\exp \left\{\bar{\pOne}
\tilde{\Omega}^{(\frac{\pOne}{\bar{\pOne}})}(p|p_S,W)\right\}
\left({\rm E}_{p}\left[
\frac{q_{Y|U}(Y|U)}{p_{Y|U}(Y|U)}
\right]\right)^{\pOne}
\nonumber\\
&=\exp \left\{\bar{\pOne}
\tilde{\Omega}^{(\frac{\pOne}{\bar{\pOne}})}
  (p|p_S,W)\right\}.
\nonumber
\end{align}
Step (a) follows from H\"older's inequality. 
Thus we have (\ref{eqn:ZsssZZZ}) in 
Lemma \ref{lm:lemmaSSz}. 
\hfill \IEEEQED 

{\it Proof of Property \ref{pr:pro1} part c):} 
We evaluate lower bounds of $F(R|p_S,W)$ to obtain 
the following chain of inequalities:
\begin{align}
& F(R|p_S,W)\MGeq{a}
\sup_{\scs  \pOne \in [0,1),
       \atop{\scs \pTwo \in [0,\frac{1}{1+2\pOne}]
            }
     }
\frac{\pTwo \pOne R -\Omega^{(\pOne,\pTwo)}(p_S,W)}
      {1+\pTwo(1+3\pOne)} 
\notag\\
&\MGeq{b}
\sup_{\scs  \pOne \in [0,1)}
\sup_{\pTwo \in [0,\frac{1}{1+2\pOne}]}
\frac{\pTwo
\Bigl\{
\pOne R -\bar{\pOne}
\tilde{\Omega}^{(\frac{\pOne}{\bar{\pOne}})}(p_S,W)
\Bigr\}
}{1+\pTwo(1+3\pOne)}
\notag\\
&\MEq{c}
\sup_{\scs  \pOne \in [0,1),
            \pThr = \frac{\pOne}{\bar{\pOne}}}
 \frac{1}{2+5\pOne}
\Bigl\{\pOne R -\bar{\pOne} 
\tilde{\Omega}^{(\frac{ \pOne }{ \bar{\pOne} })}(p_S,W)\Bigr\}
\notag\\
&\MEq{d}
\sup_{\scs \pOne= \frac{\pThr}{1+\pThr}, \pThr\geq 0
      }
\frac{\pThr R - \tilde{\Omega}^{(\pThr)}(p_S,W)}
{(1+\pThr)\left(2+ 5\frac{\pThr}{1+\pThr}\right)}
\notag\\
&=\sup_{\scs  \pThr \geq 0}\tilde{F}^{(\pThr)}(R,|p_S,W)
=\tilde{F}(R,|p_S,W).
\end{align}
Step (a) follows from the definition of ${F}(R|p_S,W)$. 
Step (b) follows from $(\ref{eqn:ZsssAZZqq})$ 
in Corollary \ref{co:CoOne}. 
Step (c) follows from that for each $\pOne \in [0,1)$,
$$
\sup_{\scs \pTwo 
\in [0,\frac{1}{1+2\pOne}]}\frac{\pTwo}{1+\pTwo[1+3\pOne]}
= \frac{1}{2+5\pOne}.
$$
Step (d) follows from that
$$
\pThr=\frac{\pOne}{\bar{\pOne}}, 
\pOne \in [0, 1) 
\Leftrightarrow
\pOne=\frac{\pThr}{1+\pThr}, \pThr \geq 0. 
$$
Thus Property \ref{pr:pro1} part c) is proved. 
\hfill\IEEEQED
}
\newcommand{\ApdaAACb}{
\subsection{
Proof of Property \ref{pr:pro1} part c), d), and e)} 
\label{sub:ApdaAACb}
In this appendix we prove Property \ref{pr:pro1} 
parts c),d), and e). 

{\it Proof of Property \ref{pr:pro1} part c):} 
\irr{
We first observe the following form of 
$\tilde{\Lambda}^{(\pThr)}(p)$:
\begin{align}
& \tilde{\Lambda}^{(\pThr)}(p)
=\sum_{\scs (u,x,y,z)
\atop{\scs \in {\cal U}\times{\cal S}
                  \times{\cal X}
                  \times{\cal Y}}}
p_{USXY}(u,s,x,y)
\notag\\
&\qquad\qquad \times 
\left[\frac{p_{Y|U}(y|u)}{p_{Y}(y)}
      \frac{p_{S}(s)}{p_{S|U}(s|u)}
\right]^{\pThr}
\label{eqn:SdZq}
\\
&=\sum_{\scs (u,s,x,y)
\atop{\scs \in {\cal U}\times{\cal S}
                  \times{\cal X}
                  \times{\cal Y}}}
p_{UY}^{\pThr}(u,y)p_{S}^{\pThr}(s)
p_{XY|SU}^{\pThr}(x,y|s,u)
\notag\\
&\quad \times 
p_{USX|Y}^{\bar{\pThr}}(u,s,x|y)p_{Y}^{1-2\pThr}(y).
\label{eqn:Sdppx}
\end{align} 
From (\ref{eqn:Sdppx}), we can see that if $\lambda \in [0,1/2]$, 
then 
\begin{align*}
& \tilde{\Lambda}^{(\lambda)}(q)
\leq |{\cal U}||{\cal S}||{\cal X}||{\cal Y}|  
\MLeq{a}|{\cal S}|^2|{\cal X}|^2 |{\cal Y}|.
\end{align*}
Step (a) follows from $p \in {\cal P}(p_S,W)$. 
We next evaluate a lower bound of $\tilde{\Lambda}^{(\pThr)}(p)$ 
for any $p \in {\cal P}(p_S,W)$, and any $\lambda \in [0,1/2]$.
Let $S_0$ be the uniformly distributed random variable 
on ${\cal S}$. For each $\lambda \in [0,1/2]$, we have the following 
chain of inequalities:
\begin{align}
& \tilde{\Lambda}^{(\pThr)}(p)
\MEq{a}{\rm E}_p
\left[ 
\left\{ 
\frac{p_{Y}(Y)}{p_{Y|U}(Y|U)}
\right\}^{-\lambda} 
\left\{
\frac{p_{S|U}(s|u)}{p_{S}(S)}
\right\}^{-\lambda}  
\right] 
\notag\\
&\MGeq{b}|{\cal S}|^{-\lambda}
{\rm E}_p
\left[ 
\left\{ 
\frac{p_{Y}(Y)}{p_{Y|U}(Y|U)}
\right\}^{-\lambda} 
\left\{
\frac{p_{S_0}(S)}{p_{S}(S)} 
\right\}^{-\lambda}  
\right] 
\notag\\
&\MGeq{c}
|{\cal S}|^{-\lambda}
\left\{ 
{\rm E}_p
\left[ 
\frac{p_{Y}(Y)}{p_{Y|U}(Y|U)}
\right] 
\right\}^{-\lambda} 
\left\{
{\rm E}_p
\left[ 
\frac{p_{S_0}(S)}{p_{S}(S)} 
\right] 
\right\}^{-\lambda}
\notag\\
&\geq  |{\cal S}|^{-\frac{1}{2}}.
\notag
\end{align}
Step (a) follows from (\ref{eqn:SdZq}).  
Step (b) follows from  
$
p_{S|U}(s|u) \leq 1
\mbox{ for }(s,u) \in {\cal S}\times {\cal U}
$
and the definition of $S_0$. 
Step (c) follows from the reverse H\"older's inequality.}
\hfill\IEEEQED

We next prove that for each $q\in {\cal P}_{\rm sh}(p_S,W)$ 
$\tilde{\Omega}^{(\lambda)}(q)$ is twice differentiable
for $\lambda \in [0,1/4]$. 

{\it Proof of Property \ref{pr:pro1} part d):} 
For simplicity of notations, set
\beqno
& & 
\underline{a} \defeq (u,s,x,y), \underline{A}\defeq (U,S,X,Y),
\underline{\cal A} \defeq 
 {\cal U} \times 
 {\cal S} \times 
 {\cal X} \times 
 {\cal Y},
\\
& & \omega^{(\mu)}_{q}(s,x,y|u) \defeq \varsigma(\underline{a}),
\tilde{\Omega}^{(\lambda)}(q)\defeq \xi(\lambda).
\eeqno

Then we have 
\beq
\tilde{\Omega}^{(\lambda)}(q)=
\xi(\lambda)=\log
\left
[\sum_{\underline{a}\in \underline{\cal A} }q(\underline{a})
{\rm e}^{\lambda {\varsigma}(\underline{a})}\right].
\label{eqn:SdfV}
\eeq
The quantity $q^{(\lambda)}(\underline{a}), \underline{a}\in {\cal A}$
has the following form:
\beq
q^{(\lambda)}(\underline{a})
={\rm e}^{-\xi(\lambda)}
q(\underline{a})
{\rm e}^{\lambda {\varsigma}(\underline{a})}.
\label{eqn:aazP011}
\eeq
By simple computations we have 
\beqa
& &\xi^{\prime}(\lambda)=
{\rm e}^{-\xi(\lambda)}
\left[\sum_{\underline{a}} q
(\underline{a})
\varsigma(\underline{a})
{\rm e}^{\lambda {\varsigma}(\underline{a})}\right],
\nonumber\\
& &\xi^{\prime\prime}(\lambda)={\rm e}^{-2\xi(\lambda)}
\nonumber\\
&&\quad \times 
\left[\sum_{\underline{a}, 
            \underline{b}\in \underline{\cal A}}
 q
(\underline{a})
 q
(\underline{b})
 \frac{\left\{{\varsigma}(\underline{a})
       -{\varsigma}(\underline{b})\right\}^2}{2}
{\rm e}^{\lambda\left\{{\varsigma}(\underline{a})
         +{\varsigma}(\underline{b})\right\}}\right]
\nonumber\\
\hspace*{-5mm}&&=
\sum_{\underline{a}, \underline{b}\in \underline{\cal A}}
q
^{(\lambda)}(\underline{a})
q
^{(\lambda)}(\underline{b})
\frac{\left\{{\varsigma}(\underline{a})-
{\varsigma}(\underline{b})\right\}^2}{2}
\nonumber\\
\hspace*{-5mm}&&=
\sum_{\underline{a}\in \underline{\cal A}}
q^{(\lambda)}(\underline{a})\varsigma^2(\underline{a})
-\left[
\sum_{\underline{a}\in \underline{\cal A}} 
q^{(\lambda)}(\underline{a})
{\varsigma}(\underline{a})\right]^2.
\label{eqn:aaz011}
\eeqa
On upper bound of $\xi^{\prime\prime}(\lambda)$,
we have the following chain of inequalities:
\begin{align}
& \xi^{\prime\prime}(\lambda)
\MLeq{a}\sum_{\underline{a}\in \underline{\cal A}}
q^{(\lambda)}(\underline{a})\varsigma^2(\underline{a})
\MEq{b}
\sum_{\underline{a}\in \underline{\cal A}}
q(\underline{a})\varsigma^2(\underline{a})
{\rm e}^{\lambda{\varsigma}(\underline{a})-\xi(\lambda)}
\nonumber\\
&={\rm e}^{-\xi(\lambda)}
\sum_{\underline{a}\in \underline{\cal A}}
q(\underline{a})\sqrt{{\rm e}^{2\lambda{\varsigma}(\underline{a})}}
\sqrt{\varsigma^4(\underline{a})}
\nonumber\\
& \MLeq{c} 
\sqrt{{\rm e}^{\xi(2\lambda)-2\xi(\lambda)}}
\sqrt{\sum_{\underline{a}\in \underline{\cal A}}
q(\underline{a})\varsigma^4(\underline{a})}.
\label{eqn:Sdppi}
\end{align}
Step (a) follows from (\ref{eqn:aaz011}).
Step (b) follows from (\ref{eqn:aazP011}).
Step (c) follows from Cauchy-Schwarz inequality and 
(\ref{eqn:SdfV}). 
Since 
$$
\sum_{\underline{a}\in \underline{\cal A}}
q(\underline{a})\varsigma^4(\underline{a})< \infty
$$
and the bound (\ref{eqn:Sdppi}), it sufficies to examine 
the quantity ${\rm e}^{\xi(2\lambda)-2\xi(\lambda)}$.
By Property \ref{pr:pro1} part c), 
this quantity is bounded for $\lambda \in [0,1/4]$.  
Hence $\xi^{\prime\prime}(\lambda)$ exists
for $\lambda \in [0,1/4]$. 
By simple analytical argument 
we know that $\xi^{\prime}(\lambda)$ exists for 
$\lambda \in [0,1/4]$. 
\hfill\IEEEQED
\newcommand{\ZsDDD}{
Since $\xi^{\prime\prime}(\lambda) \geq 0$ for 
$\lambda \in [0,1/4]$, 
$\xi^{\prime}(\lambda)$ is monotone increasing 
function of $\lambda\in [0,1/4]$. 
We also note that 
\begin{align}
\xi^{\prime}(0)
=&{\rm E}_q \left[\omega_q^{(\mu)}(S,X,Y|U)\right]
\nonumber\\
=&I_q(U;Y) -I_q(U;S)>0.
\nonumber
\end{align}
Hence, $\xi^{\prime}(\lambda)>0$ for $\lambda \in [0,1/4]$, 
implying that $\xi(\lambda)$ is a monotone
increasing function of $\lambda \in [0,1/4]$. 
}

We finally prove the part e). 

{\it Proof of Property \ref{pr:pro1} part e):} 
Fix any $(q,\lambda) \in {\cal P}(p_S,W)$ 
$\times [0,1/4]$. By the Taylor expansion of 
$\tilde{\Omega}^{(\lambda)}(q)$
with respect to $\lambda$ around $\lambda=0$,
we have that for any $(q,\lambda)\in $ 
${\cal P}(p_S,W)$ $\times [0,1/4]$ 
and for some $\gamma \in [0,\lambda]$, 
\begin{align} 
& \tilde{\Omega}^{(\lambda)}(q)
=\xi(\lambda)=\xi(0)+ \xi^{\prime}(0)\lambda
  +\frac{1}{2}\xi^{\prime\prime}(\irr{\gamma})\irr{\lambda^2}
\notag\\
&=\lambda {\rm E}_{q}\left[\tilde{\omega}_{q}(S,X,Y|U)\right]
+\frac{\irr{\lambda^2}}{2}{\rm Var}_{\irr{q^{(\gamma)}}}
\left[\tilde{\omega}_{q}(S,X,Y|U)\right]
\notag\\
&\MLeq{a} 
  \lambda C^{(\mu)}(p_S,W)
+\frac{\lambda^2}{2}{\rm Var}_{\irr{q^{(\gamma)}}}
\left[\tilde{\omega}_{q}(S,X,Y|U)\right].
\label{eqn:ZsddAA}
\end{align}
Step (a) follows from $q \in {\cal P}(p_S,W)$,  
$$
{\rm E}_{q}
\left[
\tilde{\omega}_{q}(S,X,Y|U)
\right]
= I_q(U;Y) - I_q(U;S),
$$
and the definition of $C(p_S,W)$. 
Let $(\gamma_{\rm opt},q_{\rm opt})$ 
$ \in [0, \lambda] \times {\cal P}(p_S,W)$
be a pair which attains $\rho^{(\lambda)}(p_S,W)$.
By this definition we have that
\begin{align}
&\tilde{\Omega}^{( \lambda)}(q_{\rm opt})
=\tilde{\Omega}^{( \lambda)}(p_S,W) 
\label{eqn:asWWWd}
\end{align}
and that for any $\gamma \in [0,\lambda],$
\begin{align}
&{\rm Var}_{ q_{\rm opt}^{(\gamma)}}
\left[\tilde{\omega}_{ q_{\rm opt} }(S,X,Y|U)\right]
\notag\\
&  \leq {\rm Var}_{q_{\rm opt}^{(\gamma_{\rm opt})}}
   \left[\tilde{\omega}_{q_{\rm opt}}(S,X,Y|U)\right]
=\rho^{(\lambda)}(p_S,W).
\label{eqn:asWxxd}
\end{align}
On upper bounds of $\tilde{\Omega}^{(\lambda)}(p_S,W)$, we have 
the following chain of inequalities:
\begin{align*}
&   \tilde{\Omega}^{(\lambda)}(p_S,W)
\MEq{a} 
\tilde{\Omega}^{( \lambda)}(\irr{q_{\rm opt}}) 
\nonumber\\
&\MLeq{b} \lambda C(p_S,W)
+\frac{\lambda^2}{2}
{\rm Var}_{q_{\rm opt}^{(\gamma)}}
\left[\tilde{\omega}_{q_{\rm opt}}(S,X,Y|U)\right]
\\
&\MLeq{c} \lambda {C}(p_S,W)
+\frac{\lambda^2}{2}\rho^{(\lambda)}(p_S,W)
\\
&\MLeq{d} \lambda C (p_S,W) +\frac{\lambda^2}{2}\rho(p_S,W).
\end{align*}
Step (a) follows from \irr{(\ref{eqn:asWWWd})}. 
Step (b) follows from (\ref{eqn:ZsddAA}).
Step (c) follows from \irr{(\ref{eqn:asWxxd})}. 
Step (d) follows from the definition of 
$\rho(p_S,W)$.
\hfill\IEEEQED

}
\newcommand{\ApdaAACc}{
\subsection{Proof of Property \ref{pr:pro1} part f)} 
\label{sub:ApdaAACc}

\newcommand{\UnUseTwo}{
{\it Proof of 
Property \ref{pr:pro1} part d):} 
We first observe that by simple computation we have 
\begin{align}
& \left( \frac{\rm d}{{\rm d}\lambda}
 \tilde{\Omega}^{(\lambda)}(p)\right)_{\lambda=0}
=I_p(U;Y)-I_p(U;S).
\label{eqn:axxZZ}
\end{align}
By the definition of $C(p_S,W)$, we have that  
\begin{align}
&C(p_S,W) \geq I_p(U;Y) -I_p(S;U) 
\label{eqn:aPPaZZZ} 
\end{align}
for any $p \in {\cal P}(p_S,W)$. 
Here we consider the case where $R>C(p_S,W)$. In 
this case we have $R\geq C(p_S,W)+\tau$ for some $\tau>0$. 
Applying the equality (\ref{eqn:axxZZ}) to 
$\hat{\Omega}^{(\lambda)}(p)$
$=\Omega^{(\mu,\lambda)}($$q_X,q_XW|W)$,
we have
\begin{align}
&\lim_{\lambda \to 0} 
\frac{\tilde{\Omega}^{(\lambda)}(p)}{\lambda}
=\left(\frac{\rm d}{{\rm d} \lambda } 
 \hat{\Omega}^{(\mu^*,\lambda)}(q_X|W)\right)_{\lambda=0}
\notag\\
&=I_p(U;Y)-I_p(U;S).
\label{eqn:axPss}
\end{align} 
From (\ref{eqn:axPss}), we can see that 
there exists $\lambda_0>0$ such that
\begin{align}
& \frac{\hat{\Omega}^{(\lambda_0)}(q)}{\lambda_0}
\leq I_p(U;Y)-I_p(U;S)+\frac{\tau}{2}.
\label{eqn:ZssP}
\end{align}
From (\ref{eqn:aPPaZZZ}) and (\ref{eqn:ZssP}), we have
\begin{align} 
&\frac{\tilde{\Omega}^{(\lambda_0)}(p)}{\lambda_0}
\leq C(p_S,W)+\frac{\tau}{2}.
\label{eqn:pZssPx}
\end{align}
From (\ref{eqn:pZssPx}), we have 
\beq
{\tilde{\Omega}^{(\lambda_0)}(p)}
\leq \lambda_0\left\{C(p_S,W)+\frac{\tau}{2}\right\}.
\label{eqn:AsEEE}
\eeq
Since (\ref{eqn:AsEEE}) holds 
for any $p\in {\cal P}(p_S,W)$, we have  
\beq
{\tilde{\Omega}^{(\lambda_0)}(p_S,W)}
\leq \lambda_0\left\{C(p_S,W)+\frac{\tau}{2}\right\}.
\label{eqn:AsEE}
\eeq
Then we have the following chain of inequalities:  
\begin{align*}
&\tilde{F}(R|p_S,W)
\geq \frac{\lambda_0 R -\tilde{\Omega}^{(\lambda_0)}(p_S,W)}
{2+7\lambda_0}
\MGeq{a}
\frac{\lambda_0{\tau}}{2(2+7\lambda_0)}.
\end{align*}
Step (a) follows from (\ref{eqn:AsEE}) and 
$R \geq C(p_S, W)+\tau$.
\hfill\IEEEQED
}

In this appendix we prove Property \ref{pr:pro1} part f). 
To prove this property we use the following lemma. 
\begin{lm}\label{lm:LemSaS} 
When $\tau \in (0,\rho/4]$, the maximum of 
$$
\frac{1}{2+7\lambda}
\left({\tau} \lambda-\frac{\rho}{2}\lambda^2\right)
$$
for $\lambda \in (0,1/4]$ is attained by 
the positive $\lambda_0$ satisfying 
\beq
\vartheta(\lambda_0)\defeq 
\lambda_0 + \frac{7}{4}\lambda_0^2=\frac{\tau}{\rho}.
\label{eqn:AsssD}
\eeq
Let $g(a)$ be the inverse function of $\vartheta(a)$ for $a\geq0$. 
Then the condition of (\ref{eqn:AsssD}) is equivalent 
to $\lambda_0=g(\frac{\tau}{\rho})$. The maximum is given by 
$$
\frac{1}{2+7\lambda_0}
\left(
{\tau} \lambda_0-\frac{\rho}{2}\lambda_0^2
\right)=\frac{\rho}{4}\lambda^2_0
=\frac{\rho}{4}g^2\left(\frac{\tau}{\rho}\right).
$$
\end{lm}  

By an elementary computation we can prove this lemma. 
We omit the detail.

{\it Proof of Property \ref{pr:pro1} part f):} 
When $R > {C}(p_S,W)+\tau$, 
we have the following chain of inequalities: 
\begin{align*}
&  \tilde{F}(R|p_S,W)
\geq \sup_{\lambda \in (0,1/4]} \tilde{F}^{(\lambda)}(R|p_S,W)
\\
&=\sup_{\lambda \in (0,1/4]}
\frac{\lambda R-\tilde{\Omega}^{(\lambda)}(p_S,W)}{2+7\lambda}
\\
&\MGeq{a}  
\sup_{ \lambda \in (0,1/4]} \frac{1}{2+7\lambda}
\biggl\{ \lambda \Bigl[R-C(p_S,W)\Bigr]
-\frac{\lambda^2}{2}\rho(p_S,W) \biggr\}
\\
&\MG{b} 
\sup_{\lambda  \in (0,1/4]} \frac{1}{2+7\lambda}
\left\{ \tau \lambda - \frac{\rho}{2}\lambda^2\right\}
\MEq{c}
\frac{\rho}{4} g^2 \left( \frac{\tau}{\rho}\right).
\end{align*}
Step (a) follows from Property \ref{pr:pro1} part c).
Step (b) follows from $R > {C}(p_S,W)+\tau$. 
Step (c) follows from Lemma \ref{lm:LemSaS}. 
\hfill\IEEEQED
}
Our main result is the following. 
\begin{Th}\label{Th:main}
For any $R\geq 0$, any $(p_S,W)$, and 
for any $(\varphi^{(n)},$ $\psi^{(n)})$ 
satisfying 
$(1/n)\log |{\cal K}_n| \geq R,$ 
we have 
\beqno
& &{\rm P}_{\rm c}^{(n)}(\varphi^{(n)},\psi^{(n)})
\leq 4\exp \left\{-n F(R|p_S,W)\right\}.
\eeqno
\end{Th}

It follows from Theorem \ref{Th:main} and 
Property \ref{pr:pro1} parts b) and f) that if $R$ 
is above the capacity, then the error probability 
of decoding goes to one 
exponentially and its exponent is not below $F(R|p_S,W)$.
It immediately follows from Theorem \ref{Th:main} 
that we have the following corollary. 

\begin{co}\label{co:mainCo}
For any $(p_S,W)$, we have 
\beqa
G(R|p_S,W) &\geq& F(R|p_S,W),
\label{eqn:mainIeq}
\\
{\cal R}(p_S,W) &\subseteq& \overline{\cal R}(p_S,W).
\eeqa
\end{co}

Proof of Theorem \ref{Th:main} will be given in the next section. 
The exponent function at rates outside the channel capacity 
was derived by Arimoto \cite{arimoto:73} and 
Dueck and K\"orner \cite{dueck:79}. 
The techniques used by them are not useful to prove 
Theorem \ref{Th:main}. Some novel techniques based on the 
information spectrum method introduced by Han \cite{han:book} are 
necessary to prove this theorem.


%
%

From Theorem \ref{Th:main} and Property \ref{pr:pro1} parts b) and f), 
we can obtain an explicit upper bound of 
${C}_{\rm GPC}(n,\varepsilon |p_S,W)$ 
with an asymptotically vanishing deviation from 
${C}_{\rm GPC}(p_S,W)$ $={C}(p_S,W)$. 
The strong converse theorem immediately follows 
from this corollary. We define $\nu_n=\nu_n(\varepsilon,\rho(p_S,W))$ 
by 
\begin{align}
\nu_n & \defeq 
\rho(p_S,W) \vartheta\left(
\sqrt{ \frac{4}{n\rho(p_S,W)} \log\left(\frac{4}{1-\varepsilon}\right)} 
\right)
\label{eqn:zdd}\\
&\MEq{a}
\sqrt{ 
\frac{4\rho(p_S,W)}{n}
\log\left(\frac{4}{1-\varepsilon}\right)}
+\frac{7}{n} \log \left(\frac{4}{1-\varepsilon}\right).
\nonumber
\end{align}
Step (a) follows from $\vartheta(a)=a+(7/4)a^2$.  
Since $\nu_n \to 0$ as $n\to \infty$, we have the 
smallest positive integer $n_0=n_0(\varepsilon,\rho(p_S,W))$
such that $\nu_n \leq (1/4)\rho(p_S,W)$ for $n\geq n_0$.
From Theorem \ref{Th:main}, Property \ref{pr:pro1} parts b), and f), 
we have the following corollary. 
\begin{co}\label{co:StConv}  
For each fixed $\varepsilon$ $ \in (0,1)$,  
we  choose the above positive integer 
$n_0=n_0(\varepsilon,\rho(p_S,W))$.
Then, for any $n\geq n_0$, 
\beqno
& & C_{\GPC}(n,\varepsilon|p_S,W)\leq C(p_S,W)+\nu_n. 
\eeqno
The above result together with 
\beqno
C_{\rm \GPC}(\varepsilon|p_S,W)
&=&\sup_{m \geq 1}\inf_{n \geq m}C_{\rm \GPC}(n,\varepsilon|p_S,W)
\eeqno
yields that for each $\varepsilon$ $\in (0,1)$, 
$$
{C}_{\rm GPC}(\varepsilon|p_S,W)=C(p_S,W)={C}_{\rm GPC}(p_S,W).
$$
This recovers the strong converse theorem proved 
by Tyagi and Narayan \cite{tyagi:09}. 
\end{co}

Proof of this corollary will be given in the next section.
\newcommand{\ProofCor}{  
{\it Proof of Corollary \ref{co:StConv}:}
Since $g$ is an inverse function of $\vartheta$, 
the definition (\ref{eqn:zdd}) of $\nu_n$ 
is equivalent to 
\beq
g\left( 
\frac{\nu_{n}}{\rho(p_S,W)}\right)
=\sqrt{
\frac{4}{n\rho(p_S,W)} \log\left(\frac{4}{1-\varepsilon}\right)}.
\label{eqn:zddQ}
\eeq
By the definition of $n_0=n_0(\varepsilon,\rho(p_S,W))$, 
we have that $\nu_n \leq (1/4)\rho(p_S,W)$ for $n\geq n_0$.
We assume that for $n\geq n_0$,
$R\leq {C}_{\GPC}(n,\varepsilon |p_S,W).
\label{eqn:Zassd}
$ 
Then there exists a sequence 
$\{(\varphi^{(n)},$ $\psi^{(n)})\}_{n\geq n_0}$ 
such that for $n\geq n_0$,
\beqno
& &\frac{1}{n}\log |{\cal K}_n| \geq R, 
{\rm P}_{\rm e}^{(n)}(\varphi^{(n)},\psi^{(n)})
\leq \varepsilon.
\eeqno
Then by Theorem \ref{Th:main} and Property \ref{pr:pro1} part b), 
we have that for $n\geq n_0$,
\beqa
1-\varepsilon&\leq& 
{\rm P}_{\rm c}^{(n)}(\varphi^{(n)},\psi^{(n)})
\leq 4\exp \left\{-n\tilde{F}(R|p_S,W)\right\}.\quad 
\label{eqn:Zsddd}
\eeqa
From (\ref{eqn:Zsddd}), we have that for 
$n\geq n_0$, 
\beq
\tilde{F}(R|p_S,W)\leq \frac{1}{n}
\log \frac{4}{1-\varepsilon}\MEq{a} 
\frac{\rho}{4}g^2\left(\frac{\nu_n}{\rho}\right).
\label{eqn:Zsdddii}
\eeq
Step (a) follows from (\ref{eqn:zddQ}).
Hence, by Property \ref{pr:pro1} part f),  
we have that under $\nu_n \leq (1/4)\rho(p_S,W)$,
the bound (\ref{eqn:Zsdddii}) implies 
\beq
R \leq C(p_S,W)+\nu_n.
\label{eqn:Zsdddij} 
\eeq
Since (\ref{eqn:Zsdddij}) holds for any $n \geq n_0$ 
and any $R \leq C_{\GPC}(n,\varepsilon$ $|p_S,W)$, we have
$$
C_{\GPC}(n,\varepsilon|p_S,W) \leq C(p_S,W)+\nu_n\mbox{ for }
n \geq n_0,
$$
completing the proof.
\hfill\IEEEQED  
}

\section{Proofs of the Main Results}
\label{sec:Secaa}

We first prove the following lemma.
\begin{lm}\label{lm:Ohzzz} For any $\eta>0$ and 
for any $(\varphi^{(n)},\psi^{(n)})$  
satisfying 
$(1/n)\log |{\cal K}_n| \geq R$, we have 
\begin{align}	
& {\rm P}_{\rm c}^{(n)}(\varphi^{(n)},\psi^{(n)})
\leq p_{K_nS^nX^nY^n}\hugel
\nonumber\\
& 
0\leq \frac{1}{n}
\log \frac{ p_{S}^n(S^n) W^n(Y^n|X^n,S^n)}
{Q^{\One}_{S^n}(S^n)
 Q^{\One}_{Y^n|X^nS^nK_n}(Y^n|X^n,S^n,K_n)}
\label{eqn:asXpA0}\\
& \qquad +\eta, 
\nonumber\\
& 
0\leq \frac{1}{n}
\log \frac{p_{S^n|K_n}(S^n|K_n)}
          {Q^{\Two}_{S^n|K_n}(S^n|K_n)}+\eta,
\label{eqn:asXpB0}
\\
& R\leq \frac{1}{n}\log
\frac{p_{Y^n|K_n}(Y^n|K_n)}{Q^{\Thr}_{Y^n}(Y^n)}+\eta
\huger+3{\rm e}^{-n\eta}.
\label{eqn:asXpC0}
\end{align}
In (\ref{eqn:asXpA0}), we can choose any distribution $Q^{\One}_{S^n}(S^n)$ 
on ${\cal S}^n$ and any conditional distribution 
$Q^{\One}_{Y^n|X^nS^nK_n}$ 
on ${\cal Y}^n$ given $(X^n,S^n,$ $K_n)$. 
In (\ref{eqn:asXpB0}), we can choose 
any conditional distribution $Q^{\Two}_{S^n|K_n}$ on 
${\cal S}^n$ given $K_n$.
In (\ref{eqn:asXpC0}), we can choose any conditional 
distribution $Q^{\Thr}_{Y^n}$ on ${\cal Y}^n$. 
\end{lm}

Proof of this lemma is given in Appendix \ref{sub:Apda}. 
\newcommand{\Apda}{
\subsection{
Proof of Lemma \ref{lm:Ohzzz} 
}\label{sub:Apda}


In this appendix we prove Lemma \ref{lm:Ohzzz}.
For $k\in \irr{{\cal K}_n}$, set
\beqno
& &{\cal A}_1(k)
\defeq 
\{(s^n,x^n,y^n): 
p^n_{S}(s^n)W(y^n|x^n,s^n)
\\
& &\qquad \geq Q^{\One}_{S^n}(s^n)Q^{\One}_{Y^n|X^nS^nK_n}(y^n|x^n,s^n,k)
{\rm e}^{-n\eta}\},
\\
& &{\cal A}_2(k)
\defeq \{(s^n,x^n,y^n): 
{p_{S^n|K_n}(s^n|k)}
\\
& &\qquad \geq {Q^{\Two}_{S^n|K_n}(s^n|k)}{\rm e}^{-n\eta}\}.
\eeqno
Furthermore, for $k\in {\cal K}_n$, set
\beqno
{\cal A}_3(k)
&\defeq&
\{(s^n,x^n,y^n): 
p_{Y^n|K_n}(y^n|k)
\\
& &\quad \geq |{\cal K}_n|{\rm e}^{-n\eta}
Q^{\Thr}_{Y^n}(y^n)
\},
\\
{\cal A}(k)&\defeq& \bigcap_{i=1}^3 {\cal A}_i(k).
\eeqno
Define four quantities $\Delta_i,i=0,1,2,3$ by
\begin{align*}
& \Delta_0 \defeq \Pr\{(S^n,X^n,Y^n) \in {\cal A}(K_n) \}, 
\\
& \Delta_i \defeq \Pr\{(S^n,X^n,Y^n)
\notin {\cal A}_i(K_n)\},i=1,2,
\\
& \Delta_3 \defeq 
\Pr \{(S^n,X^n,Y^n) \notin {\cal A}_3(K_n),Y^n \in {\cal D}(K_n) \}.
\end{align*}

{\it Proof of Lemma \ref{lm:Ohzzz}:} We have the following: 
\begin{align*}
&{\rm P}_{\rm c}^{(n)}=\Pr\{Y^n \in {\cal D}(K_n)\}
\leq \Pr\{(S^n,X^n,Y^n) \in {\cal A}(K_n)\}
\\
&\quad +\Pr\{(S^n,X^n,Y^n) \notin {\cal A}(K_n),Y^n \in {\cal D}(K_n)\}
\\
& \leq \sum_{i=0}^3 \Delta_i.
\end{align*}
By definition we have 
\beqa
& &\Delta_0
=p_{K_n S^nX^nY^n}\Hugel
\nonumber\\
& &\:\:\:
0\leq \frac{1}{n}
\log \frac{
p_{S}^n(S^n)
W^n(Y^n|X^nS^n)}{
Q^{\One}_{S^n}(S^n)
Q^{\One}_{Y^n|X^n S^n K_n}
         (Y^n|X^n,S^n,K_n)}+\eta,
\nonumber\\
& &\:\:\:
0\leq \frac{1}{n}
\log \frac{p_{S^n|K_n}(S^n|K_n)}
          {Q^{\Two}_{S^n|K_n}(S^n|K_n)}+\eta,
\nonumber\\
& & \left.
\frac{1}{n}\log |{\cal K}_n| \leq 
\log \frac{p_{Y^n|K_n}(Y^n|K_n)}
{Q^{\Thr}_{Y^n}(Y^n)}+\eta
\right\}.
\label{eqn:azsadaba}
\eeqa
From (\ref{eqn:azsadaba}), it follows that 
if $(\varphi^{(n)},\psi^{(n)})$  
satisfies   
$(1/n)$ $\log |{\cal K}_n| \geq R,$
then the quantity $\Delta_0$ is upper bounded by the first term 
in the right members of (\ref{eqn:asXpC0}) in Lemma \ref{lm:Ohzzz}.
Hence it suffices to show $\Delta_i\leq {\rm e}^{-n\eta}, i=1,2,3$ 
to prove Lemma \ref{lm:Ohzzz}. 
We first prove $\Delta_i \leq {\rm e}^{-n\eta}$ for $i=1,2$. 
We have the following chains of inequalities:  
\begin{align*}
& \Delta_1=\sum_{k\in {\cal K}_n }
\sum_ {\scs (s^n,x^n,y^n)
          \atop{\scs \in {\cal A}_1^{c}(k) 
          }
   } 
p_{K_nS^nX^nY^n}(k,s^n,x^n,y^n)
\\
&=\sum_{k\in {\cal K}_n }
\sum_ {\scs (s^n,x^n,y^n)
          \atop{\scs \in {\cal A}_1^{c}(k) 
          }
   }p_{S}^n(s^n)p_{X^nK_n|S^n}(x^n,k|s^n) 
\\
& \quad \times W^n(y^n|x^n,s^n)
\\
&\leq {\rm e}^{-n\eta}
\sum_{ k \in {\cal K}_n}
\sum_ {\scs (s^n,x^n,y^n)\atop{\scs \in {\cal A}_1^{c}(k)}} 
Q^{\One}_{S^n}(s^n)p_{X^nK_n|S^n}(x^n,k|s^n)
\\
&\quad \times
Q^{\One}_{Y^n|X^nS^nK_n}(y^n|x^n,s^n,k)
\leq {\rm e}^{-n\eta},
\\
&\Delta_2 =
\sum_{k\in {\cal K}_n}
\sum_ {\scs (s^n,x^n,y^n)
          \atop{\scs \in {\cal A}_2^{c}(k) 
          }
   } 
p_{K_nS^nX^nY^n}(k,s^n,x^n,y^n)
\\
&\leq {\rm e}^{-n\eta}\sum_{k\in {\cal K}_n}
p_{K_n}(k)
\sum_ {\scs (s^n,x^n,y^n)
\atop{\scs \in \tilde{\cal A}_2^{c}(k)}
   }Q^{\Two}_{S^n|K_n}(s^n|k)
\\
&\qquad \times p_{X^n Y^n|S^n K_n}(x^n,y^n|s^n,k)
 \leq {\rm e}^{-n\eta}.
\end{align*}
Finally, we prove $\Delta_3\leq {\rm e}^{-n\eta}$. 
We have the following chain of inequalities: 
\begin{align*}
& \Delta_3
= 
  \sum_{k\in {\cal K}_n}
  \sum_{\scs 
        (s^n,x^n,y^n):
         \atop{\scs 
             y^n \in {\cal D}(k), 
                \atop{\scs 
                  p_{Y^n|K_n}(y^n|k) <{\rm e}^{-n\eta}
                    \atop{\scs   
                      \times |{\cal K}_n| Q^{\Fou}_{Y^n}(y^n)
                    }
                }
         }
    }p_{K_nS^nX^nY^n}(k,s^n,x^n,y^n)
\\
&= \frac{1}{|{\cal K}_n|} 
  \sum_{k\in {\cal K}_n }
  \sum_{\scs 
         \atop{\scs 
             y^n \in {\cal D}(k), 
                \atop{\scs 
                  p_{Y^n|K_n}(y^n|k) 
                   \atop{\scs   
                  <{\rm e}^{-n\eta}|{\cal K}_n|Q^{\Fou}_{Y^n}(y^n)
                    }
                }
         }
    }p_{Y^n|K_n}(y^n|k)
\\
&\leq  {\rm e}^{-n\eta} 
  \sum_{k\in {\cal K}_n}\sum_{ y^n \in {\cal D}(k) }
Q^{\Thr}_{Y^n}(y^n)
={\rm e}^{-n\eta}Q^{\Thr}_{Y^n}
\left( \bigcup_{k \in {\cal K}_n} {\cal D}(k) \right)
\\
&\leq {\rm e}^{-n\eta}.
\end{align*}
Thus Lemma \ref{lm:Ohzzz} is proved.
\hfill\IEEEQED
}
Since $K_n$ is independent of $S^n$, we have 
\beq 
p_{S^n|K_n}(S^n|K_n)=p_{S}^n(S^n).
\label{eqn:assWW}
\eeq
From (\ref{eqn:assWW}) and Lemma \ref{lm:Ohzzz}, we immediately obtain 
the following lemma.
\begin{lm}\label{lm:Ohzzcc}
For any $\eta>0$ and for any 
$(\varphi^{(n)}, \psi^{(n)})$ satisfying  
$(1/n)\log |{\cal K}_n|\geq R,$
we have
\begin{align}
& {\rm P}_{\rm c}^{(n)}(\varphi^{(n)},\psi^{(n)})
\leq p_{K_nS^nX^nY^n}\hugel
\nonumber\\
& 
0\leq \frac{1}{n}
\log \frac{p_{S}^n(S^n) W^n(Y^n|X^n,S^n)}
{Q^{\One}_{S^n}(S^n)Q^{\One}_{Y^n|X^nS^nK_n}(Y^n|X^n,S^n,K_n)}
\:\label{eqn:asXpAz}
\\
& \qquad +\eta,
\nonumber\\
& \:R\leq \frac{1}{n}\log 
  \frac{p_{Y^n|K_n}(Y^n|K_n)       p_{S}^n(S^n)}
{Q^{\Two}_{S^n|K_n}(S^n|K_n)Q^{\Thr}_{Y^n}(Y^n)}
\label{eqn:asXpB}
\\
& \qquad +2\eta\huger + 3{\rm e}^{-n\eta}. 
\label{eqn:asXpF} 
\end{align}
In (\ref{eqn:asXpAz}), the choice of $Q^{\One}_{Y^n|X^nS^nK_n}$ 
is the same as (\ref{eqn:asXpA0}) in Lemma \ref{lm:Ohzzz}. 
In (\ref{eqn:asXpB}), we can choose any 
$Q^{\Two}_{S^n|K_n}$ and $Q^{\Thr}_{Y^n}$.
\end{lm}

For $t=1,2,$ $\cdots,n$, set 
\beqno
& {\cal U}_t \defeq {\cal K}_n \times {\cal S}^{t-1} 
   \times {\cal Y}_{t+1}^n,
   u_t \defeq (k,s^{t-1},y_{t+1}^n)\in {\cal U}_t,  
\\
& {\cal V}_t \defeq {\cal K}_n \times {\cal Y}_{t+1}^n,
   v_t \defeq (k,y_{t+1}^n) \in {\cal V}_t,
\\
& U_t \defeq (K_n,S^{t-1},Y_{t+1}^n) \in {\cal U}_t,
  V_t \defeq (K_n,Y_{t+1}^n) \in {\cal V}_t.
\eeqno
From Lemma \ref{lm:Ohzzcc}, we have the following. 
\begin{lm}\label{lm:Ohzzzca} For any $\eta>0$ and 
for any $(\varphi^{(n)},\phi^{(n)},\psi^{(n)})$  
satisfying
$(1/n)\log |{\cal K}_n| \geq R$,
we have 
\begin{align}
& {\rm P}_{\rm c}^{(n)}(\varphi^{(n)},\psi^{(n)})  
\leq 
p_{{K_n}S^nX^nY^n}\hugel
\nonumber\\
& 0\leq \frac{1}{n}
\sum_{t=1}^n
\log \frac{p_{S}(S_t)W(Y_t|X_t,S_t)}
{Q_{S_t}^{\One}(S_t)Q^{\One}_{Y_t|{X_tS_tU_t}}(Y_t|X_t,S_t,U_t)}
\nonumber\\
& \qquad +\eta,
\nonumber\\
& 
R\leq \frac{1}{n}\sum_{t=1}^n \log 
\frac{\tilde{Q}^{\Two}_{Y_t|U_t}(Y_t|U_t)
p_{S}(S_t)}
     {Q^{\Two}_{S_t|U_t}(S_t|U_t)Q^{\Thr}_{Y_t}(Y_t)}
+2\eta 
\nonumber\\
& 
\huger +3{\rm e}^{-n\eta},\qquad
\label{eqn:Zsdaa}
\end{align}
where 
for each $t=1,2,\cdots,n$, the following probability 
and conditional probability distributions:
\beq
\ba{l}
Q^{\One}_{S_t},Q^{\One}_{Y_t|X_tS_tU_t},
Q^{\Two}_{S_t|U_t},Q^{\Thr}_{Y_t}
\ea
\eeq
appearing in the first term in the right members of (\ref{eqn:Zsdaa}) 
have a property that we can choose their values arbitrary. 
In (\ref{eqn:Zsdaa}), 
$\tilde{\cal Q}^{\Two}_{Y_t|U_t}$ can be computed 
from $\{(p_{Y_i|V_i}$, 
${\cal Q}^{\Two}_{S_i|U_i}\}_{i=1}^{t-1}$ and $p_{Y_t|V_t}$, 
having the form
\begin{align*}
& \tilde{Q}^{\Two}_{Y_t|U_t}(y_t|u_t)
\\
&=
\frac{\ds \sum_{\tilde{y}^{t-1}}
\left(\prod_{i=1}^{t-1}\left\{
Q^{\Two}_{S_i|U_i}(s_i|\tilde{u}_i)
p_{Y_i|V_i}(\tilde{y}_i|\tilde{v}_i)\right\}
\right)p_{Y_t|V_t}(y_t|v_t)}
{\ds \sum_{\tilde{y}^{t}}
\left(\prod_{i=1}^{t-1}\left\{
Q^{\Two}_{S_i|U_i}(s_i|\tilde{u}_i)
p_{Y_i|V_i}(\tilde{y}_i|\tilde{v}_i)\right\}
\right)p_{Y_t|V_t}(\tilde{y}_t|v_t)},
\end{align*}
where for each $i=1,2,\cdots,t-1$, $\tilde{u}_i$ and $\tilde{v}_i$ 
are defined by
\beqno
&&\tilde{u}_i \defeq (s^{i-1},\tilde{y}_{i+1}^t,y_{t+1}^n,k)
\in {\cal S}^{i-1}\times {\cal Y}_{i+1}^{n}\times {\cal K}_n={\cal U}_i,
\\
&&\tilde{v}_i \defeq (\tilde{y}_{i+1}^t,y_{t+1}^n,k)
\in {\cal Y}_{i+1}^{n} \times {\cal K}_n={\cal V}_i.
\eeqno
\end{lm}

\begin{IEEEproof} On the probability distributions appearing in 
the right members of (\ref{eqn:asXpF}), 
we take the following choices. 
In (\ref{eqn:asXpAz}), we choose 
$Q^{\One}_{S^n}(S^n)$ so that
\beqa
Q^{\One}_{S^n}(S^n)=\prod_{t=1}^n Q^{\One}_{S_t}(S_t)
\eeqa 
and choose $Q^{\One}_{Y^n|X^nS^nK_n}$ so that 
\begin{align}
& Q^{\One}_{Y^n|X^nS^nK_n}(Y^n|X^n,S^n,K_n)
\nonumber\\
&=\prod_{t=1}^n 
Q^{\One}_{Y_t|X^{t}S^{t}Y_{t+1}^nK_n}(Y_t|X_t,S^{t},Y_{t+1}^n,K_n)
\nonumber\\
&=\prod_{t=1}^n Q^{\One}_{Y_t|X^{t}U_t}(Y_t|X_t,S_t,U_t).
\label{eqn:ddsb}
\end{align}
We define joint conditional distribution 
$Q^{\Two}_{S^nY^n|K_n}(S^n,Y^n$ $|K_n)$
on $(S^n,Y^n)$ given $K_n$ by
\begin{align}
& Q^{\Two}_{S^nY^n|K_n}(S^n,Y^n|K_n)
\nonumber\\
&=Q^{\Two}_{S^n|Y^nK_n}(S^n|Y^n,K_n)
   p_{Y^n|K_n}(Y^n|K_n)
\nonumber\\
&=\left\{\prod_{t=1}^n
Q^{\Two}_{S_t|S^{t-1}Y_{t+1}^nK_n}(S_t|S^{t-1},Y_{t+1}^n,K_n)\right\}
\nonumber\\
& \quad \times \left\{\prod_{t=1}^n
p_{Y_t|Y_{t+1}^nK_n}(Y_t|Y_{t+1}^n,K_n)\right\}.
\label{eqn:ddsbA}
\end{align}
We assume that $Q^{\Two}_{S^n|K_n}(S^n|K_n)$ 
is a marginal distribution 
of $Q^{\Two}_{S^nY^n|K_n}($$S^n,Y^n|K_n)$.
Then, in (\ref{eqn:asXpB}), we have the following:
\begin{align}
& \frac{Q^{\Two}_{Y^n|K_n}(Y^n|K_n)}
       {Q^{\Two}_{S^n|K_n}(S^n|K_n)}        
=\prod_{t=1}^n 
\frac{\tilde{Q}^{\Two}_{Y_t|S^{t-1}Y_{t+1}^nK_n}(Y_t|S^{t-1},Y_{t+1}^n,K_n)}
           {{Q}^{\Two}_{S_t|S^{t-1}Y_{t+1}^nK_n}(S_t|S^{t-1},Y_{t+1}^n,K_n)}
\nonumber\\
&=\prod_{t=1}^n 
\frac{\tilde{Q}^{\Two}_{Y_t|U_t}(Y_t|U_t)}
           {{Q}^{\Two}_{S_t|U_t}(S_t|U_t)}.
\quad \label{eqn:ddscb}
\end{align}
Based on (\ref{eqn:ddsbA}), we compute 
${\tilde{Q}^{\Two}_{S^{t-1} Y_t|Y_{t+1}^nK_n}(s^{t-1},y_t|y_{t+1}^n,k)}$
to obtain
\begin{align*}
&{\tilde{Q}^{\Two}_{S^{t-1} Y_t|Z_{t+1}^nK_n}(s^{t-1},y_t|y_{t+1}^n,k)}
\\
&=\sum_{\tilde{y}^{t-1}}
{\tilde{Q}^{\Two}_{S^{t-1}Y^{t-1}Y_t|Y_{t+1}^nK_n}
(s^{t-1},\tilde{y}^{t-1},y_t|y_{t+1}^n,k)}
\\
&=\sum_{\tilde{y}^{t-1}}
\left(\prod_{i=1}^{t-1}\left\{
Q^{\Thr}_{S_i|S^{i-1}Y_{i+1}^{t-1}Y_{t}^nK_n}(s_i|s^{i-1},
\tilde{y}_{i+1}^{t-1},y_{t}^n,k)
\right.
\right.
\nonumber\\
& \quad \left.\times 
p_{Y_i|Y_{i+1}^{t-1}Y_{t}^n K_n}
({y}_i|\tilde{y}_{i+1}^{t-1},y_{t}^n,k)
\right\}\Hugecr
p_{Y_t|Y_{t+1}^nK_n}(y_t|y_{t+1}^n,k)
\\
&=\sum_{\tilde{y}^{t-1}}
\left(\prod_{i=1}^{t-1}\left\{
Q^{\Two}_{S_i|U_i}(s_i|\tilde{u}_i)
p_{Y_i|V_i}(\tilde{y}_i|\tilde{v}_i)\right\}
\right)p_{Y_t|V_t}(y_t|v_t).
\end{align*}
Hence we have 
\begin{align*}
& \tilde{Q}^{\Two}_{Y_t|U_t}(y_t|u_t)
 =\tilde{Q}^{\Two}_{Y_t|S^{t-1}Y_{t+1}^n K_n}(y_t|s^{t-1},y_{t+1}^n,k)
\\
&=
\frac{\tilde{Q}^{\Two}_{S^{t-1}Y_t|Y_{t+1}^nK_n}(s^{t-1},y_t|y_{t+1}^n,k)}
{\tilde{Q}^{\Two}_{S^{t-1}|Y_{t+1}^nK_n}(s^{t-1}|y_{t+1}^n,k)}
\\
&=
\frac{\ds \sum_{\tilde{y}^{t-1}}
\left(\prod_{i=1}^{t-1}\left\{
Q^{\Two}_{S_i|U_i}(s_i|\tilde{u}_i)
p_{Y_i|V_i}(\tilde{y}_i|\tilde{v}_i)\right\}
\right)p_{Y_t|V_t}(y_t|v_t)}
{\ds \sum_{\tilde{y}^{t}}
\left(\prod_{i=1}^{t-1}\left\{
Q^{\Two}_{S_i|U_i}(s_i|\tilde{u}_i)
p_{Y_i|V_i}(\tilde{y}_i|\tilde{v}_i)\right\}
\right)p_{Y_t|V_t}(\tilde{y}_t|v_t)}.
\end{align*}
In (\ref{eqn:asXpB}), we choose $Q^{\Thr}_{Y^n}$ so that 
\beqa
& &Q^{\Thr}_{Y^n}(Y^n)=\prod_{t=1}^n Q^{\Thr}_{Y_t}(Y_t).
\label{eqn:ddsd}
\eeqa
From Lemma \ref{lm:Ohzzcc} and (\ref{eqn:ddsb})-(\ref{eqn:ddsd}), 
we have the bound (\ref{eqn:Zsdaa}) in Lemma \ref{lm:Ohzzzca}.
\end{IEEEproof}

For each $t=1,2,\cdots,n$, let 
$\underline{\cal Q}_t $ 
be a set of all 
\begin{align*} 
\underline{Q}_t & \defeq 
(Q_{S_t}^{\One},Q^{\One}_{Y_t|X_tS_tU_t},
Q^{\Two}_{S_t|U_tV_t},
Q^{\Thr}_{Y_t}).
\end{align*} 
Set 
\begin{align*}
\underline{\cal Q}^n & \defeq 
\prod_{t=1}^n \underline{\cal Q}_t,
\underline{Q}^n  \defeq  \left\{ \underline{Q}_t \right\}_{t=1}^n 
\in \underline{\cal Q}^n.
\end{align*}
To evaluate an upper bound of (\ref{eqn:Zsdaa}) in 
Lemma \ref{lm:Ohzzzca}. We use the following lemma, which 
is well known as the Cram\`er's bound in the large deviation principle.
\begin{lm}
\label{lm:Ohzzzb}
For any real valued random variable 
$A$ and any $\theta>0$, we have
$$
\Pr\{A \geq a \}\leq 
\exp
\left[
-\left(
\theta a -\log {\rm E}[\exp(\theta A)]
\right) 
\right].
$$
\end{lm}

Here we define a quantity which serves as an exponential
upper bound of ${\rm P}_{\rm c}^{(n)}(\varphi^{(n)},$ 
$\psi_1^{(n)},\psi_2^{(n)})$. 
Let ${\cal P}^{(n)}(p_S,W)$ be a 
set of all probability distributions 
${p}_{K_nS^nX^nY^n}$ on 
${\cal K}_n$ 
$\times {\cal S}^n$
$\times {\cal X}^n$
$\times {\cal Y}^n$
having the form:
\beqno
& &{p}_{K_nS^nX^nY^n}(k,s^n,x^n,y^n)
={p}_{K_n}(k)p_{S^n}(s^n)
\\
& &\times \prod_{t=1}^n 
{p}_{X_t|K_n X^{t-1}S^n}
(x_t|k,x^{t-1},s^n)W(y_t|x_t,s_t).
\eeqno
For simplicity of notation we use the notation $p^{(n)}$ 
for $p_{K_nS^nX^nY^n}$ $\in {\cal P}^{(n)}$
$(p_S,W)$. We assume that 
$
p_{U_tS_tX_tY_t}=p_{K_nS_{t}^nX_tY^{t}}
$
is a marginal distribution induced by $p^{(n)}$.
For $t=1,2,\cdots, n$, we simply write $p_t=$ $p_{U_tS_tX_tY_t}$. 
For $p^{(n)}$ $\in {\cal P}^{(n)}(p_S,W)$ 
and $\underline{Q}^n$ $\in \underline{\cal Q}^n$, we define 
\beqno
& &
\Omega^{(\pOne,\theta)}(p^{(n)},\underline{Q}^{n})
\defeq
\log
{\rm E}_{p^{(n)}}
\HUgebl
\HUgecl
\prod_{t=1}^n \Hugel 1
\\
&&\times \frac{p_S(S_t)W(Y_t|X_t,S_t)}
              {Q^{\One}_{S_t}(S_t)
Q^{\One}_{Y_t|X_tS_tU_t}(Y_t|X_t,S_t,U_t)}
\Huger^{\theta}\HUgecr
\\
&& \times
\left(
\prod_{t=1}^n
\left\{
\frac{
 \tilde{Q}^{\Two}_{Y_t|U_t}(Y_t|U_t)p_{S}(S_t)}
      {{Q}^{\Two}_{S_t|U_t}(S_t|U_t)
       {Q}^{\Thr}_{Y_t}(Y_t)}
\right\}^{\pOne\theta}
\right)\HUgebr.
\eeqno
Here we give a 
remark on an essential difference between 
$p^{(n)}$ $\in {\cal P}^{(n)}(p_S,W)$ and 
$\underline{Q}^n$ $\in \underline{\cal Q}^n$. 
For the former the $n$ probability distributions 
$p_t,$ $t=1,2,\cdots, n,$ are consistent with $p^{(n)}$, 
since all of them are marginal distributions 
of $p^{(n)}$. On the other hand, for the latter, $\underline{Q}^{n}$ 
is just {\it a sequence} of $n$ 
probability distributions. Hence, we may not have the 
consistency between the $n$ elements $\underline{Q}_t$, $t=1,2,\cdots,n,$ 
of $\underline{Q}^n$. 
By Lemmas \ref{lm:Ohzzzca} and \ref{lm:Ohzzzb}, we have 
the following proposition. 
\begin{pro} \label{pro:abcOhzzp}
For any 
$\pOne$, $\theta >0$, any $\underline{Q}^n \in \underline{\cal Q}^n$,
and any $(\varphi^{(n)},\psi^{(n)})$  
satisfying  
$
\frac{1}{n}\log |{\cal K}_n| \geq R,
$
we have 
\begin{align}	
& {\rm P}_{\rm c}^{(n)}(\varphi^{(n)},\psi^{(n)})
\notag\\
&
\leq 4\exp
\left\{
-n \frac{\theta R-\ds (1/n)
\Omega^{(\pOne,\theta)}
      (p^{(n)},\underline{Q}^{n})}
      {1+\theta(1+2\pOne)}
\right\}.
\label{eqn:ProOneBound}
\end{align}
\end{pro}

{\it Proof:} When 
$\theta R-\ds (1/n)\Omega^{(\pOne,\theta)}
(p^{(n)},\underline{Q}^{n})\leq 0 $, the bound (\ref{eqn:ProOneBound}) 
in Proposition \ref{pro:abcOhzzp} is obvious.
In the following argument we assume that 
$\theta R-\ds (1/n)\Omega^{(\pOne,\theta)}
(p^{(n)},\underline{Q}^{n})>0$.
We define three random variables $A_i,$$i=1,2,3$ by
\beqno
& & A_1\defeq \frac{1}{n}\sum_{t=1}^n
\log \frac{p_S(S_t)W(Y_t|X_t,S_t)}
              {Q^{\One}_{S_t}(S_t)
Q^{\One}_{Y_t|X_tS_tU_t}(Y_t|X_t,S_t,U_t)},
\\
& & A_2\defeq
\frac{1}{n}\sum_{t=1}^n \log 
\frac{
\tilde{Q}^{\Two}_{Y_t|U_t}(Y_t|U_t)
       p_{S}(S_t)}
      {{Q}^{\Two}_{S_t|U_t}(S_t|U_t)
       {Q}^{\Thr}_{Y_t}(Y_t)}.
\eeqno
Then by Lemma \ref{lm:Ohzzzca}, for any 
$(\varphi^{(n)},\psi^{(n)})$ satisfying  
$
(1/n)\log$ $|{\cal K}_n| \geq R
$
we have
\begin{align}
& {\rm P}_{\rm c}^{(n)}(\varphi^{(n)},\psi^{(n)})
\notag\\
& \leq  p_{K_nS^nX^nY^n}\{A_1\geq -\eta, A_2\geq R-2\eta\}
  +3{\rm e}^{-n\eta}
\nonumber\\
& \leq  p_{K_nS^nX^nY^n}\{A_1+\pOne A_2 
        \geq  \pOne R -\eta(1+2\pOne)\}
  +3{\rm e}^{-n\eta}
\nonumber\\
&\leq p_{K_nS^nX^nY^n}\{A \geq a \}+3{\rm e}^{-n\eta},
\label{eqn:abcawxx}
\end{align}
where we set 
$
A\defeq A_1+ \pOne A_2, a \defeq  \pOne R-\eta(1+2\pOne). 
$
Applying Lemma \ref{lm:Ohzzzb} to the first term in the right 
member of (\ref{eqn:abcawxx}), we have  
\begin{align}
& 
{\rm P}_{\rm c}^{(n)}(\varphi^{(n)},\psi^{(n)})
\nonumber\\
&\leq
\exp \left[
-\left(n\theta a -\log {\rm E}_{p^{(n)}}[\exp(n\theta A)]\right) 
\right]+3{\rm e}^{-n\eta}
\nonumber\\
&=
\exp\biggl[n\biggl\{
\theta(1+2\pOne)\eta -\theta \pOne R
\nonumber\\
& \qquad 
\left.\left.+(1/n)
\Omega^{(\pOne,\theta)}(p^{(n)},\underline{Q}^n)\right\}\right]
  +3{\rm e}^{-n\eta}.
\label{eqn:aaabv}
\end{align}
We choose $\eta$ so that 
\beq
-\eta=\theta(1+2\pOne)\eta -\theta \pOne R
  +(1/n)
\Omega^{(\pOne,\theta)}(p^{(n)},\underline{Q}^n).
\label{eqn:aaappp}
\eeq
Solving (\ref{eqn:aaappp}) with respect to $\eta$, we have 
\beqno
\eta&=&
\frac{\ds \theta \pOne R 
-(1/n)
\Omega^{(\pOne,\theta)}(p^{(n)},\underline{Q}^n)}
{1+\theta(1+2\pOne)}.
\eeqno
For this choice of $\eta$ and (\ref{eqn:aaabv}), we have
\begin{align*}
& 
{\rm P}_{\rm c}^{(n)}(\varphi^{(n)},\psi^{(n)})\leq 4{\rm e}^{-n\eta}
\\
&=4\exp
\left\{
-n\frac{\theta \pOne R- \ds (1/n) \Omega^{(\pOne,\theta)}
       (p^{(n)},\underline{Q}^{n})}
      {1+\theta(1+2\pOne)}
\right\},
\end{align*}
completing the proof. 
\hfill \IEEEQED

Set 
\beqno
\overline{\Omega}^{(\pOne,\theta)}(p_S,W)
&\defeq & 
\sup_{n\geq 1}
\max_{\scs {p}^{(n)} \in {\cal P}^{(n)}(p_S,W)}
\min_{\scs \underline{Q}^n \in \underline{\cal Q}^n}
1
\\
&& \times \frac{1}{n}\Omega^{(\pOne,\theta)}
(p^{(n)},\underline{Q}^{n}).
\eeqno
Then we have the following corollary from Proposition \ref{pro:abcOhzzp}.
\begin{co}\label{co:ProA}
For any $R\geq 0$, any $(p_S,W)$, 
any $\theta >0,\pOne \geq 0 $, and 
for any $(\varphi^{(n)},$ $\psi^{(n)})$ satisfying 
$
(1/n)\log |{\cal K}_n| \geq R,  
$
we have 
\beqno	
& & {\rm P}_{\rm c}^{(n)}(\varphi^{(n)},\psi^{(n)})
\leq 4\exp \left\{ -n
\frac{ \theta \pOne R -\overline{\Omega}^{(\pOne,\theta)}(p_S,W)}
{1+\theta(1+2\pOne)}
\right\}.
\eeqno
\end{co}

We shall call $\overline{\Omega}^{(\pOne,\theta)}(p_S,W)$ 
the communication potential. The above corollary implies that the 
analysis of $\overline{\Omega}^{(\pOne,\theta)}($ $p_S,W)$ 
leads to an establishment of a strong converse theorem for 
the state depedent channels treated in this paper. 

In the following argument we drive an explicit 
upper bound of $\overline{\Omega}^{(\pOne,\theta)}$ $(p_S,W)$. 
\newcommand{\Ft}{ {\cal F}_t } 
\newcommand{\Fi}{ {\cal F}_i } 
Set
\beqno
& &{\cal F}_t 
\defeq (p_{Y_t|V_t}, \underline{Q}_t), 
\quad {\cal F}^t \defeq \{{\cal F}_i\}_{i=1}^{t}.
\eeqno
For each $t=1,2,\cdots,n$, define a function of 
$(u_t,s_t,x_t,y_t)$
$\in {\cal U}_t$
$\times {\cal S}$ 
$\times {\cal X}$
$\times {\cal Y}$
by 
\begin{align*}
&
f_{{\cal F}^t}^{(\pOne,\theta)}
(s_t,x_t,y_t|u_t)
\nonumber\\
&\defeq  
\left\{
\frac{p_S(s_t)W(y_t|x_t,s_t)}
     {Q^{\One}_{S_t}(s_t)Q^{\One}_{Y_t|X_tS_tU_t}(y_t|x_t,s_t,u_t)}
\right\}^\theta
\\
& \quad \times
\left\{
\frac{
\tilde{Q}^{\Two}_{Y_t|U_t}(y_t|u_t)p_S(s_t)}
     {{Q}^{\Two}_{S_t|U_t}(s_t|u_t)
      {Q}^{\Thr}_{Y_t}(y_t)}
\right\}^{\pOne \theta}.
\end{align*}
Here we note that $\tilde{Q}^{\Two}_{Y_t|U_t}$ is 
uniquely determined by the component 
$(\{p_{Y_i|V_i},Q^{\Two}_{S_i|U_i}\}_{i=1}^{t-1}$ of ${\cal F}^{t-1}$ and 
$p_{Y_t|V_t}$, that is,
$$
 \tilde{Q}^{\Two}_{Y_t|U_t}
=\tilde{Q}^{\Two}_{Y_t|U_t;({\cal F}^{t-1},p_{Y_t|V_t})}.
$$
For each $t=1,2,\cdots,n$, we define a conditional probability 
distribution of $(X^t,Y^t)$ given $(K_n,S^n)$ by
\begin{align*}
& {p}_{X^tY^{t}|K_nS^n;{\cal F}^t}^{(\pOne,\theta)}
\defeq 
\left\{
p_{X^tY^{t}|K_nS^n;{\cal F}^t}^{(\pOne,\theta)}
(x^t,y^t|k,s^n)\right.
\\
& \qquad\qquad\qquad\qquad\quad 
\Bigr\}_{
(x^t,y^t,k,s^n)
   \in {\cal X}^t \times {\cal Y}^t
\times {\cal K}_n \times {\cal S}^n},
\\
& 
p_{X^tY^{t}|K_n{{S}}^n;{\cal F}^t}^{(\pOne,\theta)}
(x^t,y^t|k,s^n)
\\ 
&\defeq 
C_t^{-1}(k,s^n)
p_{X^tY^{t}|K_nS^n}(x^t,y^t|k,s^n)
\\
& \qquad \qquad \times 
\prod_{i=1}^t
f_{{\cal F}^i}^{(\pOne,\theta)}
(s_i,x_i,y_i|u_i),
\end{align*} 
where
\beqno
C_t(k,s^n)
&\defeq & 
\ba[t]{l}
\ds \sum_{x^t,y^t}p_{X^tY^{t}|K_nS^n}(x^t,y^t|k,s^n)
\vspace*{-2mm}\\
\ds \qquad \times \prod_{i=1}^t f_{{\cal F}^i }^{(\beta,\theta)}
(s_i,x_i,y_i|u_i)
\ea
\eeqno
are constants for normalization. For $t=1,2,\cdots,n$, define 
\beq
\Phi_{t, {\cal F}^t}^{(\pOne,\theta)}(k,s^n)
\defeq C_t(k,s^n)C_{t-1}^{-1}(k,s^n),
\label{eqn:defa}
\eeq
where we define $C_{0}(k,s^n)=1$ for 
$(k,s^n)\in {\cal K}_n $ $\times {\cal S}^n.$
Then we have the following lemma.
\begin{lm}\label{lm:aaa}
For each $t=1,2,\cdots,n$, and for any 
$(k,$ $s^n$ $x^t, y^t)\in {\cal K}_n$
$\times {\cal S}^n$
$\times {\cal X}^t$
$\times {\cal Y}^t$,
we have
\begin{align*}
& {p}_{X^tY^t|K_nS^n;{\cal F}^t}^{(\pOne, \theta)}
(x^t,y^t|k,s^n)
=(\Phi_{t,{\cal F}^t}^{(\pOne,\theta)}(k,s^n))^{-1}
\notag\\
& \quad \times
 p_{X^{t-1}Y^{t-1}|K_nS^n;{\cal F}^{t-1}}^{(\pOne,\theta)}
(x^{t-1},y^{t-1}|k,s^n)
\nonumber\\
& \quad \times 
f_{{\cal F}^t }^{(\pOne,\theta)}(s_t,x_t,y_t|u_t).
\end{align*}
Furthermore, we have
\begin{align}
& \Phi_{t,{\cal F}^t}^{(\pOne,\theta)}(k,s^n)
\nonumber\\
&= \sum_{x^t,y^t} 
p_{X^{t-1}Y^{t-1}|K_nS^n;{\cal F}^{t-1}}^{(\pOne,\theta)}
(x^{t-1},y^{t-1}|k,s^n)
\nonumber\\
& \quad \times p_{X_tY_t|K_nX^{t-1}S^n}
(x_t,y_t|k,x^{t-1},y^{t-1},s^n)
\nonumber\\
& \quad\times 
f_{{\cal F}_{t}}^{(\pOne,\theta)}(s_t,x_t,y_t|u_t).
\label{eqn:DfaaaK}
\end{align}
\end{lm}

Proof of this lemma is given in Appendix \ref{sub:sdfa}.
\newcommand{\Apdc}{
\subsection{Proof of Lemma \ref{lm:aaa}}\label{sub:sdfa}
In this appendix we prove Lemma \ref{lm:aaa}.  

{\it Proof of  Lemma \ref{lm:aaa}:} By the definition 
of ${p}_{X^tY^{t}|K_nS^n;{\cal F}^t}^{(\pOne,\theta)}$ 
$(x^t,y^t|k,s^n)$, for $t=1,2,\cdots,n$, we have 
\begin{align}
& p_{X^tY^{t}|K_nS^n;{\cal F}^{t}}^{(\pOne,\theta)}(x^t,y^t|k,s^n)
\nonumber\\
&=C_t^{-1}(k,s^n)
p_{X^tY^t|K_nS^n}(x^t,y^t|k,s^n) 
\nonumber\\
&\quad \times
\prod_{i=1}^t
f_{{\cal F}^i }^{(\pOne,\theta)}
(s_i,x_i,y_i|u_i,v_i).
\label{eqn:azaq}
\end{align} 
Then we have the following chain of equalities:
\begin{align} 
& p_{X^tY^{t}|K_nS^n;{\cal F}^{t}}^{(\pOne,\theta)}
(x^t,y^{t}|k,s^n)
\nonumber\\
&\MEq{a}
C_t^{-1}(k,s^n)p_{X^tY^{t}|K_nS^n}(x^t,y^t|k,s^n) 
\nonumber\\
&\quad\times \prod_{i=1}^t f_{{\cal F}^i}^{(\pOne,\theta)}
(s_i,x_i,y_i|u_i,v_i)
\nonumber\\
&=C_t^{-1}(k,{s}^n)p_{X^{t-1}Y^{t-1}|K_nS^n}(x^{t-1},y^{t-1}|k,s^n)
\nonumber\\
&\quad \times \prod_{i=1}^{t-1}
f_{{\cal F}^i}^{(\pOne,\theta)}
(s_i,x_i,y_i|u_i,v_i)
\nonumber\\
&\quad 
\times p_{ X_tY_t|X^{t-1}Y^{t-1}K_nS^n}(x_t,y_t|x^{t-1},y^{t-1},k,s^n) 
\nonumber\\
&\quad  
\times 
f_{{\cal F}^t}^{(\pOne,\theta)}(s_t,x_t,y_t|u_t,v_t)
\nonumber\\
&\MEq{b}
\frac{C_{t-1}(k,{s}^n)}{C_t(k,{s}^n)}
p_{X^{t-1}Y^{t-1}|K_nS^nv;{\cal F}^{t-1}}^{(\pOne,\theta)}
(x^{t-1},y^{t-1}|k,s^n)
\nonumber\\
&\quad 
\times p_{X_t|Y_t|X^{t-1}Y^{t-1}K_nS^n}(x_t,y_t|x^{t-1},y^{t-1},k,s^n) 
\nonumber\\
&\quad  
\times f_{{\cal F}^t }^{(\pOne,\theta)}(s_t,x_t,y_t|u_t,v_t)
\nonumber\\
&=(\Phi_{t,{\cal F}^{t}}^{(\pOne,\theta)}(k,s^n))^{-1}
\nonumber\\
&\quad \times p_{X^{t-1}Y^{t-1}|K_nS^n;{\cal F}^{t-1}}^{(\pOne,\theta)}
(x^{t-1},y^{t-1}|k,s^n)
\nonumber\\
&\quad  
\times p_{X_tY_t|X^{t-1}Y^{t-1}K_nS^n}(x_t,y_t|x^{t-1},y^{t-1},k,s^n) 
\nonumber\\
&\quad \times f_{{\cal F}^t}^{(\pOne,\theta)}(s_t,x_t,y_t|u_t,v_t).
\label{eqn:daaaq}
\end{align}
Steps (a) and (b) follow from (\ref{eqn:azaq}). 
From (\ref{eqn:daaaq}), we have 
\begin{align} 
& \Phi_{t,{\cal F}^{t}}^{(\pOne,\theta)}(k,s^n)
p_{X^tY^{t}|K_nS^n;{\cal F}^t }^{(\pOne,\theta)}
(x^t,y^t|k,s^n)
\label{eqn:daxx}
\\
&=p_{X^{t-1}Y^{t-1}|K_nS^n;{\cal F}^{t-1} }^{(\pOne,\theta)}
(x^{t-1},y^{t-1}|k,s^n)
\nonumber\\
&\quad \times
p_{X_tY_t|X^{t-1}Y^{t-1}K_nS^n}(x_t,y_t|x^{t-1},y^{t-1},k,s^n)
\nonumber\\
&\quad \times
f_{{\cal F}^t}^{(\pOne,\theta)}
  (s_t,x_t,y_t|u_t).
\label{eqn:daaxx}
\end{align}
Taking summations of (\ref{eqn:daxx}) and (\ref{eqn:daaxx}) 
with respect to $x^t,y^t$, we obtain 
\begin{align*} 
& \Phi_{t,{\cal F}^t}^{(\pOne,\theta)}(k,s^n)
=\sum_{x^t,y^t}
p_{X^{t-1}Y^{t-1}|K_nS^n;{\cal F}^{t-1} }^{(\pOne,\theta)}
(x^{t-1},y^{t-1}|k,s^n)
\nonumber\\
&\quad \times
p_{X_tY_t|X^{t-1}Y^{t-1}K_nS^n}(x_t,y_t|x^{t-1},y^{t-1},k,s^n)
\nonumber\\
&\quad \times
f_{{\cal F}^t }^{(\pOne,\theta)}
  (s_t,x_t,y_t|u_t),
\end{align*}
completing the proof.
\hfill \IEEEQED
}
Next we define a probability distribution of the random pair $(K_n,S^n)$ 
taking values in ${\cal K}_n$ $\times {\cal S}^n$ by 
\begin{align}
& p_{K_nS^n;{\cal F}^t}^{(\pOne,\theta)}(k,s^n)
\nonumber\\
&
=\tilde{C}_t^{-1} 
p_{K_nS^n}(k,s^n)
\prod_{i=1}^t 
\Phi_{i,{\cal F}^i}^{(\pOne,\theta)}(k,s^n),
\label{eqn:defzz}
\end{align}
where $\tilde{C}_t$ is a constant for normalization given by 
$$
\tilde{C}_t=\sum_{k,s^n}p_{K_nZ^n}(k,s^n)
\prod_{i=1}^t 
\Phi_{i, {\cal F}^i}^{(\pOne,\theta)}(k,s^n).
$$
By the above definition, we have
\begin{align}
\tilde{C}_n=\exp\left\{
\Omega^{(\pOne,\theta)}(p^{(n)},\underline{Q}^{n})
\right\}.
\label{eqn:Kds}
\end{align}
For $t=1,2,\cdots,n$, define 
\beq
\Lambda_{t,{\cal F}^{t}}^{(\pOne,\theta)}
\defeq \tilde{C_t}\tilde{C}_{t-1}^{-1},
\label{eqn:passdf}
\eeq
where we define $\tilde{C}_0=1$. 
Furthermore, define 
\begin{align}
& p_{K_nS_t^nX_tY^t;{\cal F}^{t-1} }^{(\pOne,\theta )}
(k,s_t^n,x_t,y^t)
\nonumber\\
&=p_{U_tS_tX_tY_t;{\cal F}^{t-1} }^{(\pOne,\theta)}
(u_t,s_t,x_t,y_t)
\nonumber\\
& \defeq  
\sum_{x^{t-1}}\sum_{\scs s^{t-1}
}
p_{K_nS^n;{\cal F}^{t-1} }^{(\pOne,\theta)}(k,s^n)
\nonumber\\
& \quad \times
p_{X^{t-1}Y^{t-1}|K_nS^n;{\cal F}^{t-1}}^{(\pOne,\theta)}
  (x^{t-1},y^{t-1}|k,s^n)
\nonumber\\
& \quad \times 
p_{X_tY_t|X^{t-1}Y^{t-1}K_nS^n}(x_t,y_t|x^{t-1},y^{t-1},k,s^n). 
\end{align}
Then, we have the following lemma, which is a key result 
to derive a single-letterized upper bound of 
$\overline{\Omega}^{(\pOne,\theta)}(p_S,W)$. 
\begin{lm}\label{lm:keylm} For any $\pOne,\theta \geq 0$, any 
$p^{(n)}\in {\cal P}^{(n)}$, and any 
$\underline{Q}^n \in \underline{\cal Q}^n$, we have 
\begin{align}
& \Omega^{(\pOne,\theta)}(p^{(n)},\underline{Q}^n)
 =\sum_{t=1}^n \log \Lambda_{t,{\cal F}^{t}}^{(\pOne,\theta)},
\label{eqn:uzapa}
\\
&\Lambda_{t,{\cal F}^t}^{(\pOne,\theta)}
= \sum_{u_t,s_t,x_t,y_t}
p_{U_tS_tX_tY_t ;{\cal F}^{t-1} }^{(\pOne,\theta)}
(u_t,s_t,x_t,y_t) 
\nonumber\\
& \qquad\qquad \times
f_{{\cal F}^t}^{(\pOne,\theta)}(s_t,x_t,y_t|u_t). 
\label{eqn:uzapaPPP}
\end{align}
\end{lm}

{\it Proof}: We first prove (\ref{eqn:uzapa}). 
We have the following: 
\begin{align}
& \exp\left\{
\Omega^{(\pOne,\theta)}(p^{(n)},\underline{Q}^{n})
\right\}
\nonumber\\
&\MEq{a}\tilde{C}_n=\prod_{t=1}^n\tilde{C}_t\tilde{C}_{t-1}^{-1}
\MEq{b}\prod_{t=1}^n
\Lambda_{t,{\cal F}^t}^{(\pOne,\theta)}. 
\label{eqn:zza}
\end{align}
Step (a) follows from (\ref{eqn:Kds}).
Step (b) follows from the definition (\ref{eqn:passdf}) of 
$\Lambda_{t,{\cal F}^t}^{(\pOne,\theta)}.$ 
From (\ref{eqn:zza}), we have (\ref{eqn:uzapa}) 
in Lemma \ref{lm:keylm}. 
We next prove (\ref{eqn:uzapaPPP}). 
Multiplying 
$\Lambda_{t,{\cal F}^t}^{(\pOne,\theta)}
=\tilde{C}_{t}/\tilde{C}_{t-1}$ to both sides of (\ref{eqn:defzz}),
we have 
\begin{align}
& \Lambda_{t,{\cal F}^t}^{(\pOne,\theta)}
 p_{K_nS^n;{\cal F}^t}^{(\pOne,\theta)}(k,s^n)
\label{eqn:aadff}
\\
&=\tilde{C}_{t-1}^{-1}
p_{K_nS^n}(k,s^n)
\prod_{i=1}^t 
\Phi_{i,{\cal F}^i}^{(\pOne,\theta)}(k,s^n)
\nonumber\\
&=
p_{K_nS^n;{\cal F}^{t-1}}^{(\pOne,\theta)}(k,s^n)
\Phi_{t,{\cal F}^t}^{(\pOne,\theta)}(k,s^n).
\label{eqn:aadffsss}
\end{align}
Taking summations of (\ref{eqn:aadff}) and (\ref{eqn:aadffsss}) 
with respect to $(k,s^n)$, we have 
\begin{align}
& 
\Lambda_{t, {\cal F}^t }^{(\pOne,\theta)}
=\sum_{k,s^n}
p_{K_nS^n;{\cal F}^{t-1} }^{(\pOne,\theta)}(k,s^n)
\Phi_{t,{\cal F}^t}^{(\pOne,\theta)}(k,s^n)
\nonumber\\
&\MEq{a}\sum_{k,s^n}\sum_{x^t,y^t} 
p_{K_nS^n;{\cal F}^{t-1}}^{(\pOne,\theta)}(k,s^n)
\nonumber\\
& \quad \times
p_{X^{t-1}Y^{t-1}|K_nS^n;{\cal F}^{t-1}}^{(\pOne,\theta)}
(x^{t-1},y^{t-1}|k,s^n)
\nonumber\\
& \quad \times 
p_{X_tY_t|X^{t-1}Y^{t-1}K_nS^n}(x_t,y_t|x^{t-1},y^{t-1},k,s^n) 
\nonumber\\
& \quad \times 
f_{{\cal F}^t}^{(\pOne,\theta)}(s_t,x_t,y_t|u_t). 
\nonumber\\
&=\sum_{\scs u_t, 
        {\scs s_t, x_t,y_t}}\sum_{x^{t-1},s^{t-1}} 
p_{K_nS^n;{\cal F}^{t-1}}^{(\pOne,\theta)}(k,s^n)
\nonumber\\
& \quad \times
p_{X^{t-1}Y^{t-1}|K_nS^n;{\cal F}^{t-1}}^{(\pOne,\theta)}
(x^{t-1},y^{t-1}|k,s^n)
\nonumber\\
& \quad \times 
p_{X_tY_t|X^{t-1}Y^{t-1}K_nS^n}(x_t,y_t|x^{t-1},y^{t-1},k,s^n) 
\nonumber\\
& \quad \times 
f_{{\cal F}^t}^{(\pOne,\theta)}(s_t,x_t,y_t|u_t). 
\label{eqn:uzapazz}
\end{align}
Step (a) follows from (\ref{eqn:DfaaaK}) in Lemma \ref{lm:aaa}.
From (\ref{eqn:uzapazz}) and the definition of 
$p_{U_tS_tX_tY_t;{\cal F}^{t-1}}^{(\pOne,\theta)}
$, we have (\ref{eqn:uzapaPPP}) in Lemma \ref{lm:keylm}. 
\hfill \IEEEQED

The following proposition is a mathematical core 
to prove our main result. 
\begin{pro}\label{pro:mainpro}
For $\theta\in (0, \pOne^{-1})$, set 
\beq
\pTwo=\frac{\theta}{1-\pOne\theta} 
\Leftrightarrow  \theta=\frac{\pTwo}{1+\pOne\pTwo}. 
\label{eqn:abaddd}
\eeq 
Then, for any positive $\pOne,$ $\pTwo$, and any 
$\theta \in (0,\pOne^{-1})$, we have 
\beqno
  \overline{\Omega}^{(\pOne,\theta)}(p_S,W)
&\leq &\frac{\Omega^{(\pOne,\pTwo)}(p_S,W)}
{1+\pOne\pTwo}.
\eeqno
\end{pro}

{\it Proof:} Set
\beqno
& &\hat{\cal Q}_n
\defeq  \{q=q_{USXY}: 
\ba[t]{l}
\pa {\cal U} \pa \leq 
   \pa {\cal K}_n \pa \pa {\cal S}^{n-1}\pa \pa {\cal Y}^{n-1}\pa\},
\ea
\\
& &\hat{\Omega}_n^{(\pOne,\theta)}(p_S,W)
\defeq
\min_{\scs 
     \atop{\scs 
     q\in \hat{\cal Q}_n
     }
}
\Omega^{(\pOne,\theta)}(q|p_S,W).
\eeqno
We recursively determine 
the sequence $\{{\cal F}^t\}_{t=1}^n$. Note 
that the component $\{p_{Y_t|V_t}\}_{t=1}^n$ 
of $\{{\cal F}^t\}_{t=1}^n$ is given. 
Hence we determine the remaining component 
$\{ \underline{Q}_t \}_{t=1}^n$.
For given ${\cal F}^{t-1}$, we choose 
$q_t=q_{U_tS_tX_tY_t}$ so that 
\begin{align}
&  q_{U_t{S}_tX_tY_t}
=p_{U_t{S}_tX_tY_t;{\cal F}^{t-1} }^{(\pOne,\theta )}
\label{eqn:Sdqw}
\end{align}
and choose the components of 
$\underline{Q}_t=(
Q^{\One}_{Y_t|X_tS_tU_t},$
$Q^{\Two}_{S_t|U_t},$ $Q^{\Thr}_{Y_t})$
such that they are the distributions 
induced by $q_{U_t}$ ${}_{S_tX_tY_t}$. Note that 
$\tilde{Q}^{\Two}_{Y_t|U_t}$ is uniquely determined 
by $({\cal F}^{t-1},$ $p_{Y_t|V_t})$. We denote 
it by $\tilde{q}_{Y_t|U_t}$. Then, for each $t=1,2,\cdots,n$, 
we have      
\begin{align}
&
f_{{\cal F}^t}^{(\pOne,\theta)}
(s_t,x_t,y_t|u_t)
\notag\\
&= 
\left\{
\frac{p_S(s_t)W(y_t|x_t,s_t)}
     {q_{S_t}(s_t)q_{Y_t|X_tS_tU_t}(y_t|x_t,s_t,u_t)}
\right\}^\theta
\notag\\
&\quad \times
\left\{
\frac{
\tilde{q}_{Y_t|U_t}(y_t|u_t)
        p_{S}(s_t)}
       {q_{S_t|U_t}(y_t|u_t)
        q_{Y_t}(y_t)}
\right\}^{\pOne\theta}.
\label{eqn:SdqwB}
\end{align}
\newcommand{\SDRR}{
appearing in
\begin{align*}
&
f_{{\cal F}^t}^{(\pOne,\theta)}
(s_t,x_t,y_t|u_t)
\\
&= 
\left\{
\frac{p_S(s_t)W(y_t|x_t,s_t)}
     {Q^{\One}_{S_t}(s_t)Q^{\One}_{Y_t|X_tS_tU_t}(y_t|x_t,s_t,u_t)}
\right\}^\theta
\\
&\quad \times
\left\{
\frac{
\tilde{Q}^{\Two}_{Y_t|U_t}(y_t|u_t)
       p_{S}(s_t)}
      {Q^{\Two}_{S_t|U_t}(y_t|u_t)
      Q^{\Thr}_{Y_t}(y_t)}
\right\}^{\pOne\theta}
\end{align*}
}
Furthermore for each $t=1,2,$ $\cdots, n$, 
we have the following chain of inequalities:
\begin{align}
& \Lambda_{t,{\cal F}^t}^{(\pOne,\theta)}
\MEq{a}
{\rm E}_{q_t}
\left[
\left\{
\frac
{p_S^{\theta}(S_t)W^{\theta}(Y_t|X_t)}
{q_S^{\theta}(S_t)q^{\theta}_{Y_t|X_t{S}_tU_t}(Y_t|X_t,{S}_t,U_t)}
\right.\right.
\nonumber\\
& \quad \times \left. \left.
\frac 
{q^{\pOne \theta}_{Y_t|U_t}(Y_t|U_t)
 p^{\pOne \theta}_{S}(S_t)}
{q^{\pOne \theta}_{S_t|U_t}(Y_t|U_t)
 q^{\pOne \theta}_{Y_t}(Y_t)}
\frac
{\tilde{q}_{Y_t|U_t}^{\pOne \theta}(Y_t|U_t)}
{      {q}_{Y_t|U_t}^{\pOne \theta}(Y_t|U_t)}
\right\} \right]
\nonumber\\
&\MLeq{b}
\left({\rm E}_{q_t}
\left[
\left\{
\frac
{p_{S}^{\theta}(S_t)W^{\theta}(Y_t|X_t,S_t)}
{q_{S_t}^{\theta}(S_t)q^{\theta}_{Y_t|X_t{S}_tU_t}(Y_t|X_t,{S}_t,U_t)}
\right.\right.\right.
\nonumber\\
& \quad \times \left.\left.\left.
\frac
{q^{\pOne \theta}_{Y_t|U_t}(Y_t|U_t)
 p^{\pOne \theta}_{S}(S_t)}
{q^{\pOne \theta}_{S_t|U_t}(S_t|U_t)
 q^{\pOne \theta}_{Y_t}(Y_t)}
 \right\}^{\frac{1}{1-\pOne \theta}}\right]
 \right)^{1-\pOne \theta}
\nonumber\\
& \quad  
\times \left(
{\rm E}_{q_t}\left[ 
\frac
{\tilde{q}_{Y_t|U_t}(Y_t|U_t)}
      {{q}_{Y_t|U_t}(Y_t|U_t)}
\right]\right)^{\pOne\theta}
\nonumber\\
&
=\exp\biggl\{\left[1- \pOne\theta \right]
\left. \Omega^{(\pOne,\frac{\theta}{1-\pOne\theta})}
(q_t|p_S,W)
\right\}
\nonumber\\
&\MEq{c}
\exp\left\{ \frac{\Omega^{(\pOne,\pTwo)}(q_t|p_S,W)}
            {1+\pOne\pTwo} \right\}
\MLeq{d}
\exp\left\{ \frac{\hat{\Omega}_n^{(\pOne,\pTwo)}(p_S,W)}
            {1+\pOne\pTwo}
\right\}
\nonumber\\
&\MEq{e}
\exp\left\{
\frac{{\Omega}^{(\pOne,\pTwo)}(p_S,W)}
{1+\pOne\pTwo}
\right\}.\:\quad\label{eqn:ssstoZ}
\end{align}
Step (a) follows from (\ref{eqn:Sdqw}), 
                      (\ref{eqn:SdqwB}), 
and Lemma \ref{lm:keylm}. 
Step (b) follows from H\"older's inequality. 
Step (c) follows from (\ref{eqn:abaddd}). 
Step (d) follows from 
$q_t \in \hat{\cal Q}_n$ and 
the definition of $\hat{\Omega}_n^{(\pOne,\pTwo)}$ 
$(p_S,W)$. Step (e) follows from that
by Property \ref{pr:pro1} part a), the bounds 
$|{\cal U}|$ $\leq |{\cal S}|$ $+|{\cal Y}|-1$ 
is sufficient to describe 
$\hat{\Omega}_n^{(\pOne,\pTwo)}(p_S,W)$.
Hence we have the following:
\begin{align}
& \min_{\scs \underline{Q}^{n}\in \underline{\cal Q}^n}
\frac{1}{n}
\Omega^{(\pOne,\theta)}
(p^{(n)},\underline{Q}^n)
\leq
\frac{1}{n}\Omega^{(\pOne,\theta)}(p^{(n)},\underline{Q}^n)
\nonumber\\
& \MEq{a}
\frac{1}{n}\sum_{t=1}^n \log \Lambda_{t,{\cal F}^t}^{(\pOne,\theta)}
 \MLeq{b}
\frac{{\Omega}^{(\pOne,\pTwo)}(p_S,W)}
{1+\pOne\pTwo}.
\qquad \label{eqn:aQ1}
\end{align}
Step (a) follows from (\ref{eqn:uzapa}) in Lemma \ref{lm:keylm}.
Step (b) follows from (\ref{eqn:ssstoZ}). 
Since (\ref{eqn:aQ1}) holds for any ${n\geq 1}$ 
and any $p^{(n)}\in {\cal P}^{(n)}$ $(p_S,W)$, we have  
$$
\overline{\Omega}^{(\pOne,\theta)}(p_S,W)
\leq 
\frac{{\Omega}^{(\pOne,\pTwo)}(p_S,W)}
{1+\pOne\pTwo},
$$
completing the proof.
\hfill \IEEEQED

{\it Proof of Theorem \ref{Th:main}: }
For any $R\geq 0$ and 
for any $(\varphi^{(n)},$ $\psi^{(n)})$ satisfying 
$
(1/n)\log |{\cal K}_n| \geq R,  
$
we have the following:
\begin{align}
& \frac{1}{n}\log
\left\{
\frac{4}{{\rm P}_{\rm c}^{(n)}
(\varphi^{(n)},\psi^{(n)})}
\right\}
\MGeq{a} 
\frac{
\theta \pOne R-\overline{\Omega}^{(\pOne,\theta)}(p_S,W)
}
{1+\theta[1+2\pOne]}
\nonumber\\
&\MGeq{b} 
\frac{\ds \frac{\pTwo \pOne R
     -\Omega^{(\pOne,\pTwo)}(p_S,W)}
             {1+\pOne\pTwo}
     }
{\ds 1+\frac{\pTwo[1+2\pOne]}{1+\pOne\pTwo}
}
=\frac{\pTwo \pOne R -\Omega^{(\pOne,\pTwo)}(p_S,W)}
{1+\pTwo(1+3\pOne)}
\nonumber\\
&=F^{(\pOne,\pTwo)}(R|p_S,W). 
\label{eqn:asff}
\end{align}
Step (a) follows from Corollary \ref{co:ProA}. Step (b) follows from 
Proposition \ref{pro:mainpro} and (\ref{eqn:abaddd}). 
Here we note that since $F^{(\pOne,\pTwo)}($ $R|p_S,W)\leq 0$ if $\pOne\pTwo=0$, 
(\ref{eqn:asff}) holds also for $\pOne\pTwo=0$. 
Since we have (\ref{eqn:asff}) for any 
nonnegative $\pOne$ and $\pTwo$, 
we have (\ref{eqn:mainIeq}) in Theorem \ref{Th:main}. 
\hfill\IEEEQED

\ProofCor 

\section{Conclusions}

We have dealt with the state dependent discrete memoryless 
channels with full state information at the sender. 
We have proved that for 
rates above the capacity the correct 
probability of decoding tends to zero exponentially 
and derived an explicit lower bound of its exponent 
function. 

\section*{\empty}
\appendix

\ApdaAAAA
\ApdaAAAb
\ApdaAACa
\ApdaAACb
\ApdaAACc
\Apda
\Apdc

\bibliographystyle{IEEEtran}
\bibliography{RefEdByOh}

\begin{thebibliography}{10}
\providecommand{\url}[1]{#1}
\csname url@samestyle\endcsname
\providecommand{\newblock}{\relax}
\providecommand{\bibinfo}[2]{#2}
\providecommand{\BIBentrySTDinterwordspacing}{\spaceskip=0pt\relax}
\providecommand{\BIBentryALTinterwordstretchfactor}{4}
\providecommand{\BIBentryALTinterwordspacing}{\spaceskip=\fontdimen2\font plus
\BIBentryALTinterwordstretchfactor\fontdimen3\font minus
  \fontdimen4\font\relax}
\providecommand{\BIBforeignlanguage}[2]{{%
\expandafter\ifx\csname l@#1\endcsname\relax
\typeout{** WARNING: IEEEtran.bst: No hyphenation pattern has been}%
\typeout{** loaded for the language `#1'. Using the pattern for}%
\typeout{** the default language instead.}%
\else
\language=\csname l@#1\endcsname
\fi
#2}}
\providecommand{\BIBdecl}{\relax}
\BIBdecl

\bibitem{gelfand:80}
S.~I. Gel'fand and M.~S. Pinsker, ``Coding for channel with random
  parameters,'' \emph{Problems of Control and Information Theory}, vol.~9,
  no.~1, pp. 19--31, 1980.

\bibitem{tyagi:09}
H.~Tyagi and P.~Narayan, ``The {G}el'fand-{P}insker channel: Strong converse
  and upper bound for the reliability function,'' in \emph{Proceedings of 2009
  IEEE International Symposium on Information Theory (ISIT)}, 2009, pp.
  1954--1957.

\bibitem{oohama2015DBCexponent}
Y.~Oohama, ``Strong converse exponent for degraded broadcast channels at rates
  outside the capacity region,'' in \emph{Proceedings of 2015 IEEE
  International Symposium on Information Theory (ISIT)}, 2015, pp. 939--943.

\bibitem{oohama2015DBCFBex}
------, ``Strong converse theorems for degraded broadcast channels with
  feedback,'' in \emph{Proceedings of 2015 IEEE International Symposium on
  Information Theory (ISIT)}, 2015, pp. 2510--2514.

\bibitem{oohama:18}
------, ``Exponential strong converse for source coding with side information
  at the decoder,'' \emph{Entropy}, vol.~20, no.~5, pp. 352 (1--32), May 2018.

\bibitem{oohama:19}
------, ``Exponential strong converse for one helper source coding problem,''
  \emph{Entropy}, vol.~21, no.~6, pp. 567 (1--36), June 2019.

\bibitem{oohama20ABCStConv}
------, ``New strong converse for asymmetric broadcast channels,'' 2020, {\it
  preprint;} available at https://arxiv.org/pdf/1604.02901.pdf, the current
  version was updated at August, 2020.

\bibitem{csiszar:78}
I.~Csisz\'ar and J.~K\"orner, ``Broadcast channels with confidential
  messages,'' \emph{IEEE Trans. Inform. Theory}, vol.~24, no.~3, pp. 339--348,
  May 1979.

\bibitem{heegard:83}
C.~Heegard and A.~El.Gamal, ``On the capacity of computer memory with
  defects,'' \emph{IEEE Trans. Inform. Theory}, vol.~29, no.~5, pp. 731--739,
  September 1983.

\bibitem{steinberg:08}
Y.~Steinberg, ``Coding for channels with rate-limited side information at the
  decoder and applications,'' \emph{IEEE Trans. Inform. Theory}, vol.~54,
  no.~9, pp. 4283--4295, September 2008.

\bibitem{oohamaItw16}
Y.~Oohama, ``Strong converse for state dependent channels with full state
  information at the sender and partial state information at the receiver,'' in
  \emph{Proceedings of 2016 IEEE Information Theory Workshop}, 2016, pp.
  126--130.

\bibitem{csiszar-korner:81}
I.~Csisz\'ar and J.~K\"orner, \emph{Information Theory, Coding Theorems for
  Discrete Memoryless Systems}.\hskip 1em plus 0.5em minus 0.4em\relax Academic
  Press, 1981.

\bibitem{tan:13}
V.~Y.~F.~Tan, ``A formula for the capacity of the general {G}el'fand-{P}insker
  channel,'' in \emph{Proceedings of 2013 IEEE International Symposium on
  Information Theory (ISIT)}, 2013, pp. 2458--2462.

\bibitem{han:book}
T.~S. Han, \emph{Information-Spectrum Methods in Information Theory}.\hskip 1em
  plus 0.5em minus 0.4em\relax Springer, 2003.

\bibitem{arimoto:73}
S.~Arimoto, ``On the converse to the coding theorem for discrete memoryless
  channels,'' \emph{IEEE Trans. Inform. Theory}, vol.~19, no.~3, pp. 357--359,
  May 1973.

\bibitem{dueck:79}
G.~Dueck and J.~K\"orner, ``Reliability function of a discrete memoryless
  channel at rates above capacity,'' \emph{IEEE Trans. Inform. Theory},
  vol.~25, no.~1, pp. 82--85, January 1979.

\end{thebibliography}

\end{document}